\documentclass[smallcondensed]{svjour3}  
\smartqed  
\usepackage{alltt                                    
	, multirow
	,subfig
	, booktabs
	, listings
	, graphicx
	,float
	,cite
	,verbatim
	,mathtools
	,url
	,amsmath
}

\usepackage[table]{xcolor}
\usepackage{xcolor, soul}
\usepackage[round]{natbib}     
\usepackage{syntax}
\usepackage{algorithmic, algorithm}
\usepackage{threeparttable}
\usepackage{framed}
\usepackage{expl3}
\usepackage{hyperref}
\usepackage{filecontents}

\usepackage{pgfplots}
\pgfplotsset{compat=1.17}
\usepackage{pgfplotstable}

\usepgfplotslibrary{statistics}

\usepackage{scalefnt}

\usepackage{tabularx}
\usepackage{lipsum}
\usepackage{array}

\usepackage{tcolorbox}
\usepackage{multirow}

\usetikzlibrary{shapes.gates.logic.US,trees,positioning,arrows,shadows}
\usetikzlibrary{shapes.multipart}
\usetikzlibrary{shapes,arrows}

\usepackage{pgf-pie}

\newif\ifpienumberinlegend
\pgfkeys{/number in legend/.code=
    \expandafter\let\expandafter\ifpienumberinlegend
    \csname if#1\endcsname
    \ifpienumberinlegend

    \def\beforenumber##1\afternumber{}%
    \fi,
    /number in legend/.default=true
}

\makeatletter
\pgfplotsset{
    boxplot/hide outliers/.code={
        \def\pgfplotsplothandlerboxplot@outlier{}%
    }
}
\makeatother

\tikzset{mynode/.style={draw=white,solid,circle,fill=green,inner sep=1pt, thick,
text=black}}
\tikzset{arrow line/.style={dashed, line width= 2.5pt, color=#1}}

\tikzstyle{vertex}=[ellipse,fill=black!25,minimum size=20pt, inner sep=0pt]
\tikzstyle{edge} = [draw,thin,-]
\tikzstyle{glabel} = [text width=1cm,text centered,font=\bf]
\pgfdeclarelayer{bg}    
\pgfsetlayers{bg,main} 

\ExplSyntaxOn
\newcommand\latinabbrev[1]{
  \peek_meaning:NTF . {
    #1\@}%
  { \peek_catcode:NTF a {
      #1., \@ }%
    {#1., \@}}}
\ExplSyntaxOff

\usepackage{verbatim}
\usetikzlibrary{arrows,shapes,backgrounds}

\tikzstyle{vertex}=[ellipse,fill=black!25,minimum size=20pt, inner sep=0pt]
\tikzstyle{edge} = [draw,thin,-]
\tikzstyle{glabel} = [text width=1cm,text centered,font=\bf]
\pgfdeclarelayer{bg}    
\pgfsetlayers{bg,main}  

\usepackage{xcolor}
\usepackage{soul}
\usepackage{pifont}
\usepackage{booktabs,makecell}
\usepackage[export]{adjustbox}
\usepackage{balance}
\usepackage{paralist}

\newcommand{\CASE}[1]{\STATE \textbf{case} #1\textbf{:} \begin{ALC@g}}
\newcommand{\ENDCASE}{\end{ALC@g}}

\newcommand{\DEFAULT}{\STATE \textbf{default:} \begin{ALC@g}}
\newcommand{\ENDDEFAULT}{\end{ALC@g}}
\newcommand{\DEFAULTLINE}[1]{\STATE \textbf{default:} }
\let\footnotesize\scriptsize

\newcounter{o}
\newsavebox{\supbox}
\newcommand{\bsup}{\begin{lrbox}{\supbox}$\tt\scriptstyle}
\newcommand{\esup}{$\end{lrbox}{}^{\usebox{\supbox}}}

\definecolor{lightpurple}{rgb}{0.8,0.8,1}
\definecolor{codebg}{RGB}{255,255,255}
\definecolor{commentcolor}{RGB}{11,140,11}
\usepackage[normalem]{ulem} 
\usepackage{xcolor}

\usepackage{ifthen}
\usepackage{amssymb}
\newboolean{showcomments}

\setboolean{showcomments}{false}

\ifthenelse{\boolean{showcomments}}
{\newcommand{\nbc}[3]{
 {\colorbox{#3}{\bfseries\sffamily\scriptsize\textcolor{white}{#1}}}
 {\textcolor{#3}{\sf\small$\blacktriangleright$\textit{#2}$\blacktriangleleft$}}
 }
}
{\newcommand{\nbc}[3]{}
 }
 
 \AtBeginDocument{%
	\paperwidth=\dimexpr
	1in + \oddsidemargin
	+ \textwidth
	+ 1in + \oddsidemargin
	\relax
	\paperheight=\dimexpr
	1in + \topmargin
	+ \headheight + \headsep
	+ \textheight
	+ 1in + \topmargin
	\relax
	\usepackage[pass]{geometry}\relax
}

\usepackage{enumitem}
\pgfplotsset{
    bar group size/.style 2 args={
        /pgf/bar shift={%
                -0.5*(#2*\pgfplotbarwidth + (#2-1)*\pgfkeysvalueof{/pgfplots/bar group skip})  + 
                (.5+#1)*\pgfplotbarwidth + #1*\pgfkeysvalueof{/pgfplots/bar group skip}},%
    },
    bar group skip/.initial=2pt,
    plot 0/.style={black,fill=black!50!white,mark=none},%
    plot 1/.style={black,fill=black!5!white,mark=none},%
    plot 2/.style={black,fill=black!80!white,mark=none},%
}

\begin{document}

\title{Automatic Prediction of Rejected Edits in Stack Overflow}

\author{Saikat Mondal         \and
        Gias Uddin \and
        Chanchal Roy 
}


\institute{Saikat Mondal \at
          Software Research Lab, Department of Computer Science \\ University of Saskatchewan, Canada \\
          \email{saikat.mondal@usask.ca}           
           \and
           Gias Uddin \at
           Data Intensive Software Analytics (DISA) Lab, Department of Electrical and Software Engineering \\
           University of Calgary, Canada \\ 
           \email{gias.uddin@ucalgary.ca}
           \and
           Chanchal Roy \at
           Software Research Lab, Department of Computer Science \\ University of Saskatchewan, Canada \\
           \email{chanchal.roy@usask.ca}
}

\date{Received: date / Accepted: date}

\maketitle

\begin{abstract}

The content quality of shared knowledge in Stack Overflow (SO) is crucial in supporting software developers with their programming problems. Thus, SO allows its users to suggest edits to improve the quality of a post (i.e., question and answer). However, existing research shows that many suggested edits in SO are rejected due to undesired contents/formats or violating edit guidelines. Such a scenario frustrates or demotivates users who would like to conduct good-quality edits. Therefore, our research focuses on assisting SO users by offering them suggestions on how to improve their editing of posts. 
First, we manually investigate 764 (382 questions + 382 answers) rejected edits by rollbacks and produce a catalog of 19 rejection reasons. 
Second, we extract 15 texts and user-based features to capture those rejection reasons. 
Third, we develop four machine learning models using those features. Our best-performing model can predict rejected edits with 69.1\% precision, 71.2\% recall, 70.1\% F1-score, and 69.8\% overall accuracy. 
Fourth, we introduce an online tool named \texttt{EditEx} that works with the SO edit system. \texttt{EditEx} can assist users while editing posts by suggesting the potential causes of rejections.
We recruit 20 participants to assess the effectiveness of \texttt{EditEx}. Half of the participants (i.e., treatment group) use \texttt{EditEx} and another half (i.e., control group) use the SO standard edit system to edit posts.
According to our experiment, \texttt{EditEx} can support SO standard edit system to prevent 49\% of rejected edits, including the commonly rejected ones. However, it can prevent 12\% rejections even in free-form regular edits. 
The treatment group finds the potential rejection reasons identified by \texttt{EditEx} \emph{influential}. Furthermore, the median workload suggesting edits using \texttt{EditEx} is \emph{half} compared to the SO edit system.
\end{abstract}


\section{Introduction}\label{sec:introduction}

The adoption, growth, and continued success of an online question and answering (Q\&A) site such as Stack Overflow (SO) depend on two major factors -- (1) participation of users and (2) quality of the shared knowledge~\citep{Bagozzi-ParticipationLinuxUserGroups-JMC2006,Lakhani-FreeUserToUserAssistance-JRP2003,Parnin-CrowdDoc-TechReport2012}. 
SO thus introduces an edit system to promote quality by allowing its users to communicate on the quality of the posts through editing. In particular, collaborative editing helps to keep posts clear, relevant, and up-to-date.
For example, users often edit posts to fix grammar and spelling mistakes, clarify the meaning, and add related resources or hyperlinks. Unfortunately, many suggested edits in SO get rejected because of undesired (i.e., it does not satisfy the post owner) editing or violating edit guidelines \citep{footnote22}. However, edits can be rejected in two ways -- \emph{rollback} and \emph{expert review}. Rollback reverts a post to a previous version in the edit history \citep{footnote39} and thereby rejects one/multiple revisions. On the other hand, experts (e.g., users with a reputation score $\geq$ $2K$) can reject if suggested edits do not improve the quality of the posts.
However, manual identification of undesired edits or edits that violate the editing guidelines wastes community time and effort. For example, one user responded to the issue of the identification of undesired edits manually, \emph{``It takes time to read and parse through those questions when I am trying to spend my time more efficiently reading through the actual question and figuring out how to answer it appropriately''} \citep{footnote31}. 
At least $921$ users supported this comment by casting upvotes. It suggests that manually identifying undesired edits wastes users' valuable time and resources. On the other hand, users who suggest edits and later get rejected become frustrated because many users (especially novices) are unaware of editing guidelines \citep{mondal2021rollback}. Unfortunately, the existing editing system of SO does not identify the rejected edits with the potential rejection reasons. Therefore, a study on automatic identification of rejected edits with reasons is warranted to assist SO users.

Realizing the need for an automated tool, some users started writing personal scripts to identify undesired edits programmatically. For example, one user wrote a script to automatically identify greetings (e.g., hello, dear) while reviewing suggested edits \citep{footnote31}. 
Such a scenario urges a system that identifies the potential rejected edits. However, capturing all rejection reasons using simple rule-based scripts is challenging. Thus, a robust technique (e.g., machine learning classifiers) needs to be introduced to reasonably identify rejected edits and the potential reasons behind those rejections.
Wang et al.~\citep{Wang-SOEdit-TSE2018} investigate the rejected edits in SO. They analyze 369 rejected edits (by rollbacks) of answers and identify 12 reasons (e.g., undesired text formatting). Their study shows the empirical evidence of the complexity and diversity of reasons that can contribute to the rejection of suggested edits. However, we are unaware of any existing edit assistance system that automatically identifies rejected edits with reasons to support the current editing system of SO.

\begin{figure}[t]
	\centering
	\includegraphics[width=4.7in]{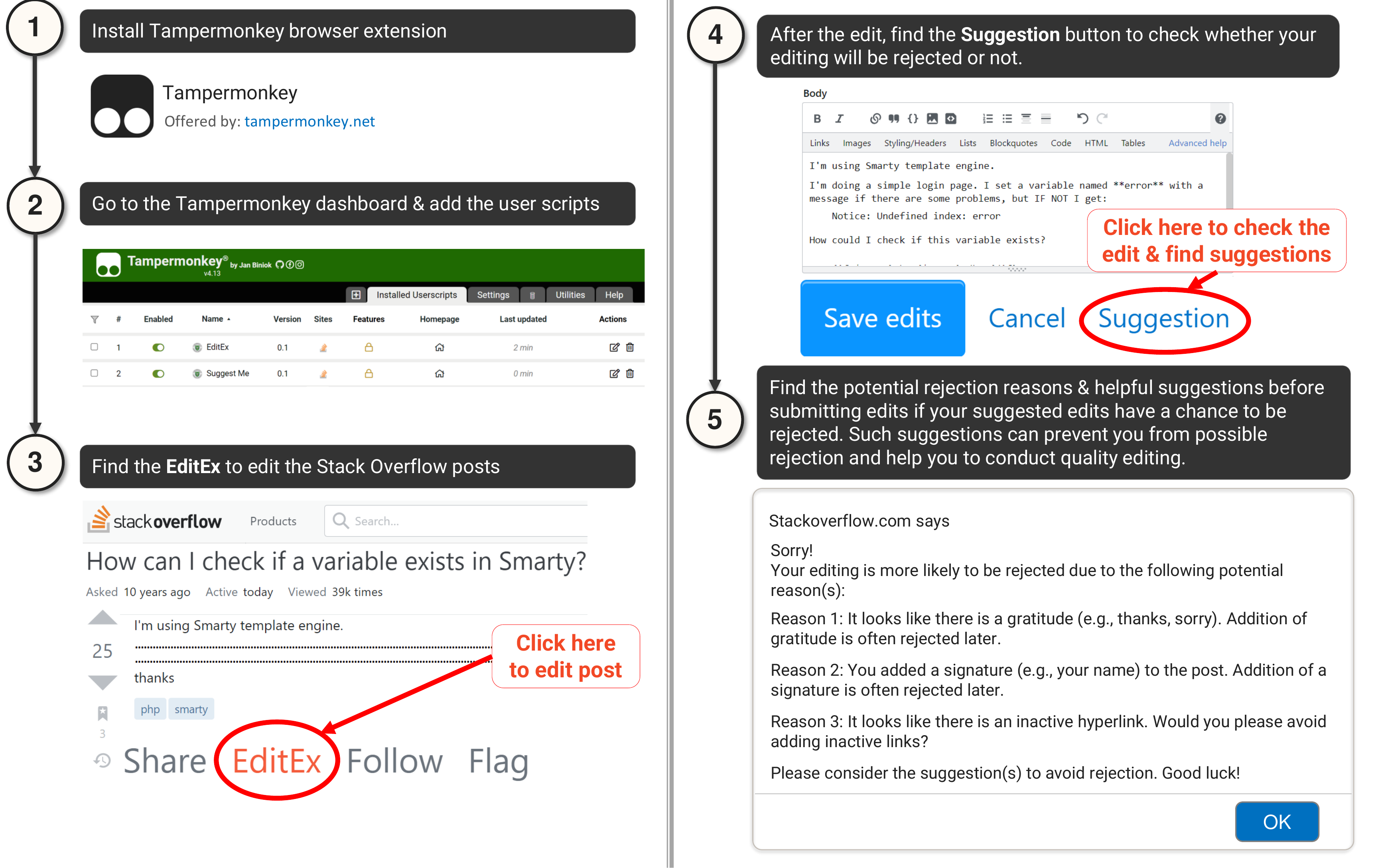}
	\caption{An overview of the \texttt{EditEx} workflow.}
	\label{fig:editex-workflow}
\end{figure}

This study focuses on assisting SO users by offering them automated suggestions on how to improve their editing of posts. 
First, we manually analyzed 764 rejected edits (382 questions + 382 answers). We identified 19 rejection reasons (Table \ref{table:list-of-rollback-reasons}), seven of which were not reported by Wang et al.~\citep{Wang-SOEdit-TSE2018}. 
Second, we extract 15 texts and user-based features to capture those rejection reasons. 
Third, using those features, we develop four machine learning classifiers (e.g., random forest). According to the experiment, our best-performing model can predict rejected edits with 69.1\% precision, 71.2\% recall, 70.1\% F1-score, and 69.8\% accuracy. 
Fourth, we introduce an online tool named \texttt{EditEx} that works with the SO edit system. \texttt{EditEx} can assist users while editing posts by identifying the rejected edits and the potential causes of rejections.

Fig. \ref{fig:editex-workflow} shows the overview of \texttt{EditEx} workflow. The workflow is described as follows.
%
 (1) On the client-side, users need to install Tampermonkey\footnote{\url{https://www.tampermonkey.net}}, one of the most popular userscript managers. It offers an effortless way to manage userscripts. In addition, it is available as a browser extension for all the popular browsers, such as Chrome, Firefox, Safari, Microsoft Edge, and Opera. 
(2) Users require to add two JavaScript scripts for the \emph{EditEx} and \emph{Suggestion} interface with the SO edit system. Tampermonkey enables users to add JavaScript scripts that can be used to modify web pages. 
(3) \emph{EditEx} enables users to edit the posts.
(4) \emph{Suggestion} captures the texts (before \& after edits) and user information (e.g., reputation score, name). It then transmits texts and user information to the server where the machine learning classifiers have been deployed. The server-side application extracts the features. It then predicts whether the edit will be rejected or not and identify the potential reasons (if rejected) using the machine learning classifiers.
(5) Finally, it shows the decision (rejected/accepted) of the classifier and suggests to users with the potential reasons if rejected.

We recruited 20 participants and divided them into treatment and control groups. The treatment group used \texttt{EditEx} and the control group used the SO standard edit system to edit posts. We survey the participants after completing their edits. According to survey results, the treatment group found the potential reasons identified by \texttt{EditEx} behind the rejections \emph{influential}. Moreover, 49\% rejections (including the commonly rejected ones) were prevented upon following the suggestions offered by \texttt{EditEx}.
However, \texttt{EditEx} is also capable of preventing 12\% rejections even in \emph{free-form} regular edits. Free-form edits refer to those edits that are not related to any specific rejection reasons. This tool also significantly reduces the effort of suggesting edits and makes participants more confident.
The following major stakeholders in crowd-sourced knowledge-sharing platforms that use collaborative editing features can be benefited from the findings from our study and the tool \texttt{EditEx}: \begin{inparaenum}
\item forum like SO designers to improve the edit system, 
\item forum users to assist their edit behavior, and 
\item software engineering researchers to study and improve collaborative editing support in crowd-shared platforms.
\end{inparaenum}

\smallskip\noindent\textbf{Data Availability.} The datasets analyzed during this study are available in our online appendix \citep{replicationPackage}.

\smallskip\noindent\textbf{Deviations} of this study from our registered report \citep{mondal2020rollbackRegisteredProtocol} are discussed in Section \ref{sec:deviation}.

\smallskip\noindent\textbf{Structure of the Article.} 
The rest of this article is structured as follows. Section \ref{sec:rejection-catalog} represents a catalog of edit rejection reasons. In Section \ref{sec:rejected-edit-prediction-model}, we discuss a model that predicts rejected edits with potential reasons. Section \ref{sec:editex-tool} discusses our online tool named \texttt{EditEx}, its architecture, and analyzes its effectiveness. Feature ranking, the reasons behind the misclassifications of our model, and the implications of our study are discussed in Section \ref{sec:discussion}. Section \ref{sec:threat} focuses on the threats to validity, Section \ref{sec:relatedwork} discusses the related work and finally, Section \ref{sec:conclusion} concludes our study.

\section{A Catalog of Edit Rejection Reasons}
\label{sec:rejection-catalog}

SO introduces an edit system to improve the quality of the posts. However, edits may not always be satisfactory and thus can be rejected. Wang et al.~\citep{Wang-SOEdit-TSE2018} conduct an initial investigation on rejected edits of answers by rollback and expose 12 potential reasons behind rejections. We extend their investigation by manually investigating rejected edits by rollbacks of questions and answers. This section discusses our manual investigation process first and later summarizes the identified edit rejection reasons.

\subsection{Dataset Preparation}
\label{subsec:data-collection}

We downloaded the September 2019 (which was the start time of the journey of this study) data dump of SO from the Stack Exchange site \citep{datadumpapi}. The data dump stores the history of all events (e.g., rollback and edit body) of the posts. In this study, we only investigate the revisions where users made edits to the body of posts. Our data dump contains a total of 1,116,473 rejected edits (72,159 questions + 44,314 answers) by rollbacks and 26,604,779 accepted edits (13,624,495 questions + 12,980,284 answers). We manually analyze rejected edits to explore the rejection reasons. However, accepted edits will be used later to train and test our machine learning classifiers in Section \ref{subsec:rejectedEditPredictor}. We randomly sampled a statistically significant sample size from rejected and accepted edits. 

To achieve a confidence level of 95\% with a confidence interval of 5\% \citep{boslaugh2012statistics}, we randomly sampled -- (1) 382 from 72,159 rollback edits of questions, (2) 382 from 44,314 rollback edits of answers, (3) 385 from 13,624,495 accepted edits of questions, and (4) 385 from 12,980,284 accepted edits of answers. We use the following formula to compute the size of our random sample.

\begin{equation}
    \frac{Nz^2p(1-p)}{e^2N+z^2p(1-p)}
\end{equation}

\noindent where N is the population size (e.g., 72,159), z is the Z-score corresponding to a particular confidence level (e.g., 1.96 for a confidence level of 95\%), e is the confidence interval (e.g., 5\%), and p is population proportion (e.g., 0.5) \citep{Wang-SOEdit-TSE2018}.

The data dump only stores the latest contents of the questions and answers after each revision. However, we need both the texts before and after rollback/acceptance to analyze which suggested edits are rejected/accepted. Therefore, we collect the \texttt{PostId} and \texttt{RevisionGUID} for each of our randomly selected revisions from the data dump. We manually save the web pages using \texttt{PostId} from the SO site containing each post's revision history. Next, we find the target revision from the history using the \texttt{RevisionGUID}. Note that \texttt{RevisionGUID} is a unique ID used to find a particular revision. Each revision contains the text before and after rollback/acceptance. Finally, we extract those target texts.

\begin{table}[!t]
	\centering
	\captionsetup{justification=centering, labelsep=newline}
	\caption{Summary of the \emph{nineteen} manually derived reasons behind rollback edits}
	\label{table:list-of-rollback-reasons}
	\resizebox{4.8in}{!}{%
	\rowcolors{1}{}{lightgray!30}
    \begin{tabular}{p{4.8cm}|p{7.8cm}|c|c|c|c} \toprule
    
    \rowcolor{lightgray!0} \multicolumn{1}{p{4.8cm}|}{\textbf{Rollback Reasons}} &  \multicolumn{1}{c|}{\textbf{Description}} & \rotatebox[origin=l]{90}{\textbf{Question}} & \rotatebox[origin=l]{90}{\textbf{Answer}} & \rotatebox[origin=l]{90}{\textbf{Wang et al.}} & \rotatebox[origin=l]{90}{\textbf{Count (\%)}} \\ \midrule
    
    \multirow{5}{*}{\textbf{Undesired Text Formatting}}           & Users change the format of the texts unnecessarily. Such changes include changing the font, text cases (uppercase/lowercase), emphasizing text by making them bold/italic, adding or removing spaces/newlines, creating bullet/number list, and formatting text term as code element or vice versa. & \multirow{5}{*}{\ding{51}} & \multirow{5}{*}{\ding{55}} & \multirow{5}{*}{\ding{51}} & \multirow{5}{*}{157 (20.5\%)}\\ 
    
    \multirow{2}{*}{\textbf{Undesired Text Add/Remove}}           & Users add texts with less or no impact on the quality/clarification of posts or remove essential texts. & \multirow{2}{*}{\ding{51}} & \multirow{2}{*}{\ding{51}} & \multirow{2}{*}{\ding{51}} & \multirow{2}{*}{223 (29.2\%)} \\ 
    
    \multirow{5}{*}{\textbf{Undesired Text Change}}  & Users make undesired changes of the sentence structures (e.g., simple, complex), tenses (e.g., present, past), voices (e.g., active, passive), rewording, interchanging contractions by root words, acronyms/abbreviations by elaborations and vice versa.  & \multirow{5}{*}{\ding{51}} & \multirow{5}{*}{\ding{51}} & \multirow{5}{*}{\ding{51}} & \multirow{5}{*}{85 (11.1\%)} \\ 
    
    \multirow{4}{*}{\textbf{Incorrect Text Change}}               & Users perform rewording with incorrect terms, grammatical and spelling mistakes, incorrect changes in software versions or specifications, changes that alternate the meaning of the sentence. & \multirow{4}{*}{\ding{51}} & \multirow{4}{*}{\ding{51}} & \multirow{4}{*}{\ding{51}} & \multirow{4}{*}{86 (11.3\%)} \\ 
    
    \multirow{4}{*}{\textbf{Undesired Code Formatting}}           & Users make the undesired modification of code indentation (e.g., addition/removal of spaces/newlines), addition/removal of line numbers, split/merge of code segments, changes in text cases (e.g., \emph{select} to \emph{SELECT} in a SQL query). & \multirow{4}{*}{\ding{51}} & \multirow{4}{*}{\ding{51}} & \multirow{4}{*}{\ding{51}} & \multirow{4}{*}{74 (9.7\%)} \\ 
    
    \multirow{2}{*}{\textbf{Undesired Code Add/Remove}}           & Unwanted code statements (e.g., alternative solutions) are added, or essential segments are removed. & \multirow{2}{*}{\ding{51}} & \multirow{2}{*}{\ding{51}} & \multirow{2}{*}{\ding{51}} & \multirow{2}{*}{89 (11.6\%)}\\ 
    
    \multirow{3}{*}{\textbf{Undesired Code Change}}              & Users make undesired code changes, such as changing options of a command, changing APIs, refactoring (e.g., variable renaming), and editing comments. & \multirow{3}{*}{\ding{51}} & \multirow{3}{*}{\ding{51}} & \multirow{3}{*}{\ding{51}} & \multirow{3}{*}{24 (3.1\%)} \\ 
    
    \multirow{3}{*}{\textbf{Incorrect Code Change}}               & Users make incorrect changes to the data type of variables, function return types, function arguments, arithmetic expressions. & \multirow{3}{*}{\ding{51}} & \multirow{3}{*}{\ding{51}} & \multirow{3}{*}{\ding{51}} & \multirow{3}{*}{58 (7.6\%)} \\ 
    
    \multirow{3}{*}{\textbf{Status Update}}                       & Users add/remove personal notes to clarify confusion, append messages missed during the submission of their posts, and acknowledge users’ responses. & \multirow{3}{*}{\ding{51}} & \multirow{3}{*}{\ding{51}} & \multirow{3}{*}{\ding{55}} & \multirow{3}{*}{77 (10.1\%)} \\ 
    
    \multirow{2}{*}{\textbf{Emotion Add/Remove}}                  & Addition/removal of words/sentences/emoticons that represent personal emotion. & \multirow{2}{*}{\ding{51}} & \multirow{2}{*}{\ding{51}} & \multirow{2}{*}{\ding{51}} & \multirow{2}{*}{6 (0.8\%)} \\ 
    
    \multirow{2}{*}{\textbf{Gratitude Add/Remove}}                  & Addition/removal of thanksgiving sentences (e.g., thank you, cheers!). & \multirow{2}{*}{\ding{51}} & \multirow{2}{*}{\ding{51}} & \multirow{2}{*}{\ding{55}} & \multirow{2}{*}{42 (5.5\%)} \\ 
    
    
    \multirow{1}{*}{\textbf{Greetings Add/Remove}}                & Addition/removal of greeting/salutations (e.g., hello, dear). & \multirow{1}{*}{\ding{51}} & \multirow{1}{*}{\ding{55}} & \multirow{1}{*}{\ding{55}}  & \multirow{1}{*}{4 (0.5\%)} \\  
    
    \multirow{3}{*}{\textbf{Undesired Ref. Modification}}    & Users add inactive hyperlinks, inappropriate images, or diagrams with posts. On the contrary, sometimes they remove essential hyperlinks/images or unreasonably modify them. & \multirow{3}{*}{\ding{51}} & \multirow{3}{*}{\ding{51}} & \multirow{3}{*}{\ding{51}} & \multirow{3}{*}{85 (11.1\%)} \\ 
    
    \multirow{2}{*}{\textbf{Signature Add/Remove}}               & Addition/removal of users' names, hyperlinks to personal websites. & \ding{51} & \multirow{2}{*}{\ding{51}} & \multirow{2}{*}{\ding{55}} & \multirow{2}{*}{14 (1.8\%)} \\ 
    
    \multirow{3}{*}{\textbf{Partial Acceptance}}                  & Revision is rolled back, but part of the changes are still accepted. Then, the accepted changes are included in later revisions. & \multirow{3}{*}{\ding{51}} & \multirow{3}{*}{\ding{51}} & \multirow{3}{*}{\ding{51}} & \multirow{3}{*}{4 (0.5\%)} \\ 
    
    \multirow{2}{*}{\textbf{Deprecation Note Add/Remove}}                    & Addition/removal of deprecation notes inside the body of an answer. & \multirow{2}{*}{\ding{55}} & \multirow{2}{*}{\ding{51}} & \multirow{2}{*}{\ding{55}} & \multirow{2}{*}{1 (0.1\%)} \\ 
    
    \multirow{2}{*}{\textbf{Duplication Note Add/Remove}}                    & Addition/removal of duplication notes inside the body of a question. & \multirow{2}{*}{\ding{51}} & \multirow{2}{*}{\ding{55}} & \multirow{2}{*}{\ding{55}} & \multirow{2}{*}{11 (1.4\%)} \\ 
    
    \multirow{4}{*}{\textbf{Community Trust}} & The reputation score estimates how much the community trusts a user. Users with low reputations often do not follow the guidelines when editing posts. Thus, their edits are rejected more than the highly reputed users. & \multirow{4}{*}{\ding{51}} & \multirow{4}{*}{\ding{51}} & \multirow{4}{*}{\ding{55}} & \multirow{4}{*}{N/A} \\
    
    
    \multirow{5}{*}{\textbf{Other}}                               &Other reasons include asking questions inside the answer, adding solutions inside the question, interchanging the position of texts, introducing spam (i.e., users deface text/code to promote a product or service, insert garbage texts), and changing posts completely. & \multirow{5}{*}{\ding{51}} & \multirow{5}{*}{\ding{51}} &  \multirow{5}{*}{\ding{51}} & \multirow{5}{*}{40 (5.2\%)} \\ \bottomrule
    \end{tabular}
    }
\end{table}

\subsection{Edit Rejection Reasons}\label{sec:edit-rejection-reasons}

Table \ref{table:list-of-rollback-reasons} summarizes the rollback edit reasons.
We (two authors of this paper) manually investigate the randomly selected 764 (382 questions + 382 answers) rejected edits by rollbacks. We consider the rollback reasons identified by Wang et al.~\citep{Wang-SOEdit-TSE2018} as the baseline during our analysis. However, we discuss the rollback edit reasons in multiple interactive sessions. We then analyze 200 rollback edits (100 questions + 100 answers) from our selected dataset and label the reasons. For a given rollback edit, we meticulously analyze the texts before and after rollback to see the edits that cause a rollback. Our in-depth investigation exposes a total of \emph{nineteen} potential reasons. \emph{Twelve} of them were identified by Wang et al., and the remaining \emph{seven} are new reasons. The new reasons are -- (1) status update, (2) gratitude add/remove, (3) greetings add/remove, (4) signature add/remove, (5) deprecation note add/remove, (6) duplication note add/remove, and (7) community trust. We then measure the agreement using Cohen's Kappa \citep{cohen1968weighted, cohen1960coefficient}. The value of {\large $\kappa$} was $0.98$, which means the strength of the agreement is almost perfect. Next, we resolve the remaining few disagreements by discussion. However, the agreement level indicates that any coder can do the rest of the labeling without introducing individual bias. Thus, the first author of this paper analyzes the remaining dataset and manually labels the reasons.



\section{A Model to Predict Potential Rejection of Suggested Edits}
\label{sec:rejected-edit-prediction-model}

The identification of undesired editing that causes rejection is essential to promote quality editing. However, manual differentiation of undesired and accepted edits can waste a lot of time and effort of users. Thus, this study aims to assist SO users by offering them automated support while editing a post. Specifically, we attempt to exploit cues in the textual contents of suggested edits to build a classifier that can automatically determine the rejected edits with potential rejection reasons. In this section, we answer the research question as follows.

\begin{tcolorbox}[colframe=black!50,colback=white,left=0pt,right=1pt,top=1pt,bottom=1pt, boxrule=1pt,arc=1pt, width=4.7in, center]
    \textbf{$RQ_1$)} To what extent do our classifiers predict the rejected edits with the potential reasons?
\end{tcolorbox}

\smallskip\noindent For \textbf{$RQ_1$}, we define the following null hypothesis.

\begin{tcolorbox}[colframe=black!50,colback=white,left=0pt,right=1pt,top=1pt,bottom=1pt, boxrule=1pt,arc=1pt, width=4.7in, center]
    \emph{$H_1$:} The accuracy of our developed classifiers is not better than a random classifier with 50\% accuracy.
\end{tcolorbox}

\begin{figure}[h]
	\centering
	\includegraphics[width=4.2in]{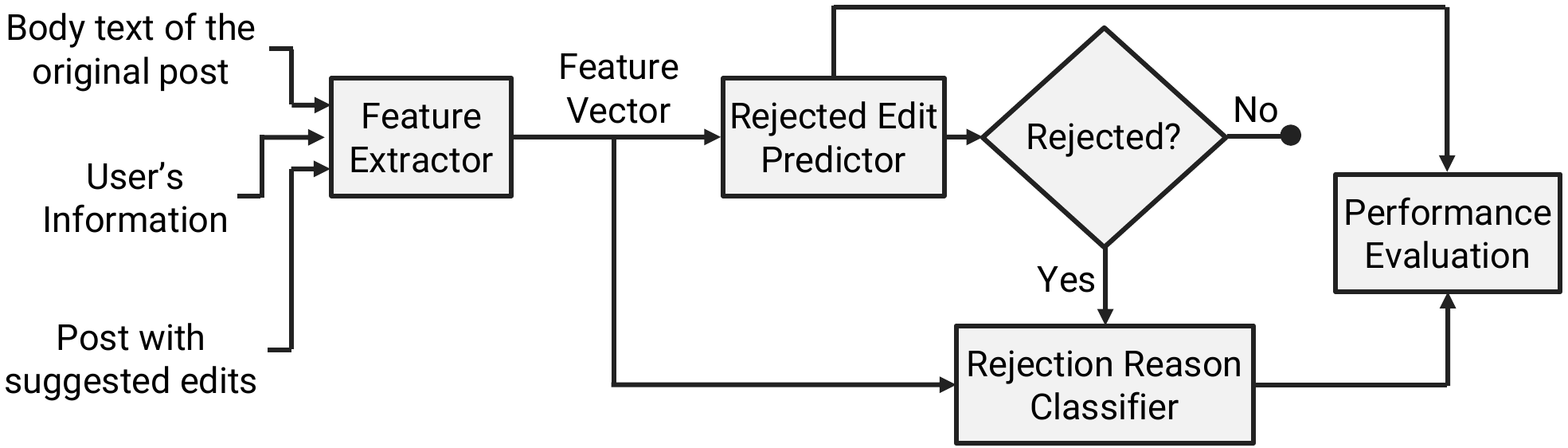}
	\caption{Workflow to predict rejected edits of SO with reasons.}
	\label{fig:flowchart-rq-1}
\end{figure}

Fig. \ref{fig:flowchart-rq-1} shows the workflow of how we can detect the potential rejected edits of SO with reasons. Our prediction pipeline includes three main components: \emph{Feature Extractor}, \emph{Rejected Edit Predictor} and \emph{Rejection Reason Classifier}. First, the feature extractor takes the body text of the original post, the post with suggested edits, and the user's information (e.g., reputation) as inputs. Then, it produces a feature vector based on the predictor variables. Second, the rejected edit predictor takes the feature vectors as input and outputs a dichotomous variable, \texttt{rejected}. The value of \texttt{rejected} is $1$ if the predictor determines that the suggested edit will most likely be rejected, and it is $0$ otherwise. Third, the rejection reason classifier takes the corresponding feature vector and texts as inputs if the value of \texttt{rejected} is $1$. Then, it outputs the potential reasons for rejection. Finally, we measure the performances of the rejected edit predictor and rejection reason classifier.

\begin{table}[!t]
	\centering
	\captionsetup{justification=centering, labelsep=newline}
	\caption{Features of our predictor}
	\label{table:features-of-predictor}
	\resizebox{4in}{!}{%
    \begin{tabular}{l|l|p{5cm}} \toprule
    \textbf{Predictor}  & \textbf{Features}       & \textbf{Covered Reasons} \\ \midrule
    
    \multirow{6}{*}{\textbf{Editing Distance}}  & Text/Code Formatting &  Undesired Text/Code Formatting \\ \cmidrule{2-3}
                                                & \multirow{2}{*}{Text/Code Modification} &  Undesired Text/Code Add/Remove, Undesired Text/Code Change \\ \cmidrule{2-3}
                                                & Deface Post      &  Other \\ \cmidrule{2-3}
                                                & Complete Change of Post &  Other \\ \midrule
    
    \multirow{1}{*}{\textbf{Status}}            & Status & Status Update  \\ \midrule
    
    \multirow{1}{*}{\textbf{Gratitude}}         & Gratitude & Gratitude Add/Remove  \\ \midrule
    
    \multirow{1}{*}{\textbf{Greetings}}         & Greetings & Greetings Add/Remove  \\ \midrule
    
    \multirow{2}{*}{\textbf{Reference}}         & Reference Modification & Undesired Reference Modification  \\ \cmidrule{2-3}
                                                & Inactive Hyperlink & Undesired Reference Modification  \\ \midrule
    
    \multirow{1}{*}{\textbf{Signature}}         & Signature & Signature Add/Remove  \\ \midrule
     
    \multirow{1}{*}{\textbf{Deprecation}}       & Deprecation Note & Deprecation Note Add/Remove  \\ \midrule
    
    \multirow{1}{*}{\textbf{Duplication}}       & Duplication Note & Duplication Note Add/Remove  \\ \midrule
    
    \multirow{1}{*}{\textbf{Reputation}}        & Reputation Score & Community Trust  \\ \bottomrule
    
    \end{tabular}
    }
\end{table}

\subsection{Feature Extractor}
\label{subsec:featureExtraction}

Table \ref{table:features-of-predictor} summarizes the features employed to predict whether suggested edits will be rejected or not. We extracted \emph{fifteen} texts \& user-based features to predict the potential rejected edits. Each feature is connected to one or multiple reasons behind a rejected edit. 
Note that we discarded \emph{emotion}, which was included in our registered report \citep{mondal2020rollbackRegisteredProtocol}, and added \emph{reputation}. The reasons behind such decisions are discussed in Section \ref{sec:deviation}.
However, this section discusses how we extract the features as follows.



\smallskip\noindent\textbf{Text/Code Formatting.} 
Text or code formatting refers to the changes in their presentation styles. For example, consider the text with HTML tags --  $<$p$>$ I am using $<$b$>$ C\# $<$/b$>$ programming language $<$/p$>$. Here, C\# is formatted as bold. Someone can reject the bold format of C\# by removing the $<$b$>$...$<$/b$>$ tags. However, in both cases, the content remains unchanged. The extracted content will be ``I am using C\# programming language''. We thus detect the text/code formatting in the following ways.

\begin{tcolorbox}[colframe=black!10, colback=black!2, left=1pt,right=1pt,top=1pt,bottom=1pt,arc=0mm, title = {Determine Text Formatting}, boxrule=0.0pt, fonttitle={\color{black}}]

    $textFormat =$ \emph{T\textsubscript{br}} != null $\&\&$ \emph{T\textsubscript{ar}} != null 
    
    \hspace{6mm} $\&\&$ \emph{D\textsubscript{e}}(\emph{T\textsubscript{br}}, \emph{T\textsubscript{ar}}) $==$ 0 
    
    \hspace{12mm} $\&\&$ \emph{D\textsubscript{e}} (\emph{T\textsubscript{brwt}},\emph{T\textsubscript{arwt}}) $!=$ 0

\end{tcolorbox}


    
    
    
    


\noindent where \emph{T\textsubscript{brwt}}: Texts before rollback with HTML tags (e.g., $<$p$>$ I am using $<$b$>$ C\# $<$/b$>$ programming language $<$/p$>$), 
\emph{T\textsubscript{br}}: Texts before rollback (e.g., I am using C\# programming language),
\emph{T\textsubscript{arwt}}: Texts after rollback with HTML tags (e.g., $<$p$>$ I am using C\# programming language $<$/p$>$),
\emph{T\textsubscript{ar}}: Texts after rollback (e.g., I am using C\# programming language), and 
\emph{D\textsubscript{e}}: \emph{Levenshtein} \citep{yujian2007normalized, editDistance} editing distance. However, we remove new lines, strip leading \& trailing spaces from \emph{T\textsubscript{br}} \& \emph{T\textsubscript{ar}}, and make them lowercase before processing.

\begin{tcolorbox}[colframe=black!10, colback=black!2, left=1pt,right=1pt,top=1pt,bottom=1pt,arc=0mm, title = {Determine Code Formatting}, boxrule=0.0pt, fonttitle={\color{black}}]

    $codeFormat =$ \emph{C\textsubscript{br}} != null $\&\&$ \emph{C\textsubscript{ar}} != null 
    
    \hspace{6mm} $\&\&$ \emph{D\textsubscript{e}}(\emph{C\textsubscript{br}}, \emph{C\textsubscript{ar}}) $==$ 0 
    
    \hspace{12mm} $\&\&$ \emph{D\textsubscript{e}} (\emph{C\textsubscript{brwt}},\emph{C\textsubscript{arwt}}) $!=$ 0

\end{tcolorbox}


    
    
    
    


\noindent where \emph{C\textsubscript{brwt}}: Code before rollback with HTML tags, 
\emph{C\textsubscript{br}}: Code before rollback,
\emph{C\textsubscript{arwt}}: Code after rollback with HTML tags, and 
\emph{T\textsubscript{ar}}: Code after rollback. Like text formatting, we remove new lines, strip leading \& trailing spaces from \emph{C\textsubscript{br}} \& \emph{C\textsubscript{ar}}, and make them lowercase before processing.



\smallskip\noindent\textbf{Text/Code Modification.}
Text/code modification refers to text/code addition/removal or change of existing text/code.



\begin{tcolorbox}[colframe=black!10, colback=black!2, left=1pt,right=1pt,top=1pt,bottom=1pt,arc=0mm, title = {Determine Text Modification}, boxrule=0.0pt, fonttitle={\color{black}}]

    if (\emph{T\textsubscript{br}} != null $\&\&$ \emph{T\textsubscript{ar}} != null $\&\&$ \emph{D\textsubscript{e}}(\emph{T\textsubscript{br}}, \emph{T\textsubscript{ar}}) $!=$ 0)$\{$
    
    
    \hspace{6mm} $textModification =$ \emph{D\textsubscript{e}}(\emph{T\textsubscript{br}}, \emph{T\textsubscript{ar}})
    
    $\}$
    else 
    
    \hspace{6mm} $textModification = False$

\end{tcolorbox}


    
    
    
    


\noindent We normalize \emph{textModification} with respect to the character length of \emph{T\textsubscript{br}} since \emph{T\textsubscript{br}} varies. 
We determine the code modification in the same way as text modification.









\smallskip\noindent\textbf{Deface Post.}
Remove the texts or code entirely.

\begin{tcolorbox}[colframe=black!10, colback=black!2, left=1pt,right=1pt,top=1pt,bottom=1pt,arc=0mm, title = {Deface Text}, boxrule=0.0pt, fonttitle={\color{black}}]

$defaceText =$ (\emph{T\textsubscript{br}} != null $\&\&$ \emph{T\textsubscript{ar}} == null) $||$ (\emph{T\textsubscript{ar}} != null $\&\&$ \emph{T\textsubscript{br}} == null)

\end{tcolorbox}







\noindent We determine the deface code in the same way as the deface text. However, deface post will be \emph{True} if \emph{defaceText} or \emph{defaceCode} becomes \emph{True}.

\smallskip\noindent\textbf{Complete Change of Post.}
Users change texts or code segments entirely.

\begin{tcolorbox}[colframe=black!10, colback=black!2, left=1pt,right=1pt,top=1pt,bottom=1pt,arc=0mm, title = {Determine Complete Change of Text}, boxrule=0.0pt, fonttitle={\color{black}}]

$completeChangeText =$ \emph{T\textsubscript{br}} != null $\&\&$ \emph{T\textsubscript{ar}} != null

\hspace{6mm} $\&\&$ (\emph{D\textsubscript{e}}(\emph{T\textsubscript{br}}, \emph{T\textsubscript{ar}}) $==$  Len(\emph{T\textsubscript{br}}) $||$ \emph{D\textsubscript{e}}(\emph{T\textsubscript{br}}, \emph{T\textsubscript{ar}}) $==$ Len(\emph{T\textsubscript{ar}}))

\end{tcolorbox}








\noindent We determine the complete code change in the same way as the text change. However, complete change of post will be \emph{True} if \emph{completeChangeText} or \emph{completeChangeCode} becomes \emph{True}.

\smallskip\noindent\textbf{Status.}
To detect the addition/deletion of status, we attempt to find a keyword match to \emph{T\textsubscript{br}}/\emph{T\textsubscript{ar}} from a keywords list, \emph{L\textsubscript{kw}}. Here, \emph{L\textsubscript{kw}} = $\{$\emph{edit, update, note, ps}$\}$ \citep{mondal2021rollback}.

\begin{tcolorbox}[colframe=black!10, colback=black!2, left=1pt,right=1pt,top=1pt,bottom=1pt,arc=0mm, title = {Determine Status}, boxrule=0.0pt, fonttitle={\color{black}}]

    \emph{status = False}
    
    for (\emph{keyword} in \emph{L\textsubscript{kw}})$\{$
    
    \hspace{6mm} if ((Match(\emph{keyword}, \emph{T\textsubscript{br}}) == $True$ $\&\&$ Match(\emph{keyword}, \emph{T\textsubscript{ar}}) == $False$) $||$
    
    \hspace{6mm} (Match(\emph{keyword}, \emph{T\textsubscript{ar}}) == $True$ $\&\&$ Match(\emph{keyword}, \emph{T\textsubscript{br}}) == $False$))$\{$
    
    \hspace{12mm} \emph{status = True}
    
    \hspace{12mm} return \emph{status}
    
    \hspace{6mm} $\}$
    
    
    
    
    
    $\}$
    
    \emph{return status}

\end{tcolorbox}

\smallskip\noindent\textbf{Gratitude.}
We detect the addition/removal of gratitude similarly to status. However, here the keyword list, \emph{L\textsubscript{kw}} = $\{$\emph{welcome, thanks, sorry, appreciated, thank, ty} (i.e., thank you), \emph{thx, regards, tia} (i.e., thanks in advance)$\}$ \citep{mondal2021rollback}.

\smallskip\noindent\textbf{Greeting.}
We detect the addition/removal of greeting similarly to status. However, here the keyword list, \emph{L\textsubscript{kw}} = $\{$\emph{hi, hello, hey, dear, greetings, hai, guys, hii, howdy, hiya, hay, heya, hola, hihi, salutations}$\}$ \citep{footnote31}.

\smallskip\noindent\textbf{Reference Modification.}
To detect reference modification, we extract the values of the \emph{href} attribute of $<$a$>$ tag from \emph{T\textsubscript{brwt}} \& \emph{T\textsubscript{arwt}}.
Then insert the hyperlinks into two lists -- \emph{LST\textsubscript{br}}: list of hyperlinks found from texts before rollback, and \emph{LST\textsubscript{ar}}: list of hyperlinks found from texts after rollback.

\begin{tcolorbox}[colframe=black!10, colback=black!2, left=1pt,right=1pt,top=1pt,bottom=1pt,arc=0mm, title = {Determine Reference Modification}, boxrule=0.0pt, fonttitle={\color{black}}]

    \emph{referenceModification = False}
    
    
    for (\emph{reference} in \emph{LST\textsubscript{br}})$\{$
    
    \hspace{6mm} if (Contain(\emph{reference}, \emph{LST\textsubscript{ar}}) == $False$)$\{$
    
    \hspace{12mm} \emph{referenceModification = True}
    
    \hspace{12mm} return \emph{referenceModification}
    
    \hspace{6mm} $\}$

    $\}$
    
    for (\emph{reference} in \emph{LST\textsubscript{ar}})$\{$
    
    \hspace{6mm} if (Contain(\emph{reference}, \emph{LST\textsubscript{br}}) == $False$)$\{$
    
    \hspace{12mm} \emph{referenceModification = True}
    
    \hspace{12mm} return \emph{referenceModification}
    
    \hspace{6mm} $\}$

    $\}$
    
    \emph{return status}

    
    
    

\end{tcolorbox}











\smallskip\noindent\textbf{Inactive Hyperlink.}
To detect the inactive (e.g., broken/dead) hyperlink, we check the \texttt{HTTP} response of each of the hyperlinks of \emph{LST\textsubscript{ar}}. We decide whether a hyperlink is inactive or not based on the response code.


\smallskip\noindent\textbf{Signature.}
To detect the addition/removal of the signature, we extract and store the full name, first part, and last part (if any) of two users (who suggested edit \& who rolled back) into a list, \emph{L\textsubscript{name}}. We then detect the addition/removal of the signature similarly to status.







\smallskip\noindent\textbf{Deprecation Note.}
We detect the addition/removal of deprecation notes similarly to status. However, here the keyword list, \emph{L\textsubscript{kw}} = $\{$\emph{deprecation, deprecate, old code}$\}$ \citep{mondal2021rollback}.




\smallskip\noindent\textbf{Duplication Note.}
We detect the addition/removal of duplication notes similarly to status. However, here the keyword list, \emph{L\textsubscript{kw}} = $\{$\emph{duplicate, duplication}$\}$ \citep{mondal2021rollback}.





\smallskip\noindent\textbf{Reputation Score.}
We compute the reputation score of users to estimate -- (1) how much the community trusts them and (2) whether they follow the editing guidelines. 
The official data dump of SO only reports the latest reputation scores of the users, which are not appropriate for our analysis. We thus use the snapshot of user activities (e.g., votes, acceptances, bounties) to compute the reputation score of users during their editing of posts. In particular, we use a standard equation provided by the SO to calculate the reputation score \citep{HDR-StackOverflow-2020}.


\subsection{Rejected Edit Predictor}
\label{subsec:rejectedEditPredictor}

In this section, we first describe machine learning classifiers (Section \ref{subsec:machineLearningModel}) and their evaluation setup (Section \ref{subsec:modelEvaluationSetup}). Then we evaluate the performance of our classifiers in identifying rejected edits (Section \ref{subsec:performanceEvaluation}). Finally, we construct the baseline models and report their performance in Section \ref{subsec:baselineModel}.

\subsubsection{Machine Learning Models}
\label{subsec:machineLearningModel}

The relationship between edit categories (rejected/accepted) and their corresponding feature values might be complex. Thus, we choose the following four popular machine learning classification techniques with different learning strategies to identify the potential rejected edits. They are widely used in the relevant studies \citep{TUT-Saha-2013,UAC-Ponzanelli-2014,AII-Rahman-2015, beyer2018automatically}.

\smallskip\noindent\textbf{Decision Trees (DT)} is a non-parametric supervised machine learning technique for classification and regression. Non-parametric means it does not make any assumptions about the underlying data distribution. The intuition behind decision trees is that simple decision rules are inferred from the dataset features and continually split the training set until all data points belonging to each class are isolated. In particular, this technique employs different heuristics (e.g., entropy, information gain) to decide which feature to be used for the subsequent split of the training set. The commonly used decision trees are \texttt{ID3}, \texttt{C4.5}, and \texttt{CART}. However, \texttt{ID3} can only be used when features are categorical. \texttt{C4.5} and \texttt{CART} are extensions of \texttt{ID3}, which can work with features of both categorical and continuous data. Here, we use \texttt{CART} since our extracted features have both continuous and categorical values.

\smallskip\noindent\textbf{Random Forest (RF)} 
is a supervised machine learning technique. The `forest' it builds is an ensemble of many decision trees, usually trained with the `bagging' method. The underlying principle behind the ensemble model is that a group of weak learners come together to form a strong learner. Ensemble learners thus improve the performance of single classifiers by inducing several classifiers and combining them to obtain a new classifier that outperforms every one of them \citep{EBS-Polikar-2006}. RF is scalable to any number of dimensions and usually has acceptable performance. However, it adds additional randomness to the model while growing the trees. For example, instead of searching for the most important feature while splitting nodes, it searches for the best feature among a random subset of features. RF thus prevents the overfitting of datasets by creating random subsets of the features.

\smallskip\noindent\textbf{K-Nearest Neighbors(KNN)}
is a non-parametric method employed in classification and regression problems \citep{NCA-Goldberger-2005}. It does not use the training data points to perform any generalization. Thus, the KNN's training phase is much faster than other classification algorithms. In KNN, K represents the number of nearest neighbors, which is the core factor in deciding a data point's label (i.e., class). This technique finds the K closest neighbors of a target point using distance measures (e.g., Euclidean distance). Then, each neighbor votes for their class, and the class with the most votes is taken as the prediction.

\smallskip\noindent\textbf{eXtreme Gradient Boosting (XGBoost)} is a scalable tree boosting technique that predicts a target variable by combining an ensemble of estimates from a set of more simplistic and weaker models \citep{chen2016xgboost}. It is a supervised learning algorithm and can be employed for both classification and regression. XGBoost is an extension to gradient boosted decision trees (GBM) with improved speed and performance. However, it is faster than other algorithms because of its parallel and distributed computing. XGBoost performs well because of its robust handling of various data types, relationships, distributions, and a variety of hyperparameters. In addition, XGBoost has inbuilt cross-validation and a variety of regularizations, which helps reduce overfitting.

\subsubsection{Model Evaluation Setup}
\label{subsec:modelEvaluationSetup}

\smallskip\noindent\textbf{Dataset Selection.} We used the randomly sampled dataset (rejected \& accepted) from Section \ref{subsec:data-collection} to train and test our machine learning classifiers. However, we keep the training and testing data separate since we have one time-dependent feature (i.e., reputation score). For example, consider a user suggested edits to multiple posts at different times. The reputation score of users increases over time. Thus, the reputation score of that user will be lower while suggesting prior edits than the later edits. Therefore, we use earlier edits as the training set and later edits as the test set to ensure that past data is not predicted based on future data. In particular, we take 70\% samples that were edited relatively earlier to train the machine learning classifiers and use the remaining 30\% to test them.

\begin{table}[!htb]
	\centering
	\captionsetup{justification=centering, labelsep=newline}
	\caption{Confusion matrix}
	\label{table:confusion-matrix}
	\resizebox{3in}{!}{%
    \begin{tabular}{c|c|c}
    \multicolumn{1}{c|}{\textbf{Sample Size (N)}}                                                   & \multicolumn{1}{c|}{\begin{tabular}[c]{@{}c@{}}\textbf{Expected}\\ (\emph{Rejected})\end{tabular}} & \multicolumn{1}{c}{\begin{tabular}[c]{@{}c@{}}\textbf{Expected}\\ (\emph{Accepted})\end{tabular}}         \\ \midrule
    \multicolumn{1}{c|}{\begin{tabular}[c]{@{}c@{}}\textbf{Predicted}\\ (\emph{Rejected})\end{tabular}} & True Positive (\textbf{TP}) & False Positive (\textbf{FP}) \\ \midrule
    \multicolumn{1}{c|}{\begin{tabular}[c]{@{}c@{}}\textbf{Predicted}\\ (\emph{Accepted})\end{tabular}}  & False Negative (\textbf{FN}) & True Negative (\textbf{TN}) \\ \midrule
    \end{tabular}
    }
    \vspace{-4mm}
\end{table}

\smallskip\noindent\textbf{Performance Metrics.} The selection of the evaluation criteria is vital to guarantee a reliable assessment of the prediction models. In a binary classification problem like rejected edit prediction, a confusion matrix (e.g., Table \ref{table:confusion-matrix}) records the correctly and incorrectly recognized examples of each class. Therefore, we can obtain several metrics from the given confusion matrix to independently evaluate models' performance for both positive and negative classes. 

The machine learning community often measures the \emph{classification accuracy} as a simple scalar performance metric for binary classification. Classification accuracy measures the ratio of correctly classified edits into rejected \& accepted classes with respect to all classified edits. However, according to \citet{LFI-He-2008}, accuracy might lead to an incorrect conclusion as the measure is highly sensitive to changes in data. In such cases, \emph{precision} is a useful metric to capture the effect on a classifier performance of having a larger number of negative examples \citep{TRB-Davis-2006}. In particular, precision measures the ratio of correctly classified edits into a class (i.e., rejected/accepted) with respect to all edits classified into that class. However, \citet{LFI-He-2008} argued that precision is still sensitive to changes in the data distribution, and it cannot assert how many positive examples are classified incorrectly. 
Unlike precision, \emph{recall} is not sensitive to data distribution that measures the ratio of correctly classified edits with respect to the actually observed edits as true instances. However, any assessment based solely on recall would be inadequate, as it provides no insight into how many examples are incorrectly classified as positives. 
Therefore, neither precision nor recall can provide a reliable assessment of classification performance \citep{AEA-Calefato-2019}. However, these individual scalar metrics can be combined to build more reliable classification performance measures. Specifically, these aggregated performance metrics include the F-measure that represents the harmonic mean of precision and recall. We thus measure precision, recall, F1-score, and overall accuracy to conclude the models' performance better. They can be measured as follows.

    
    \hspace{6mm} \emph{Precision} $= \frac{TP}{TP+FP}$ \hspace{20mm} \emph{Recall} $= \frac{TP}{TP+FN}$ 
    
    \hspace{6mm} \emph{F1-Score} $\frac{2 \times Precision \times Recall}{Precision + Recall}$ \hspace{8mm} \emph{Accuracy} $= \frac{TP+TN}{TP+FP+TN+FN}$
    



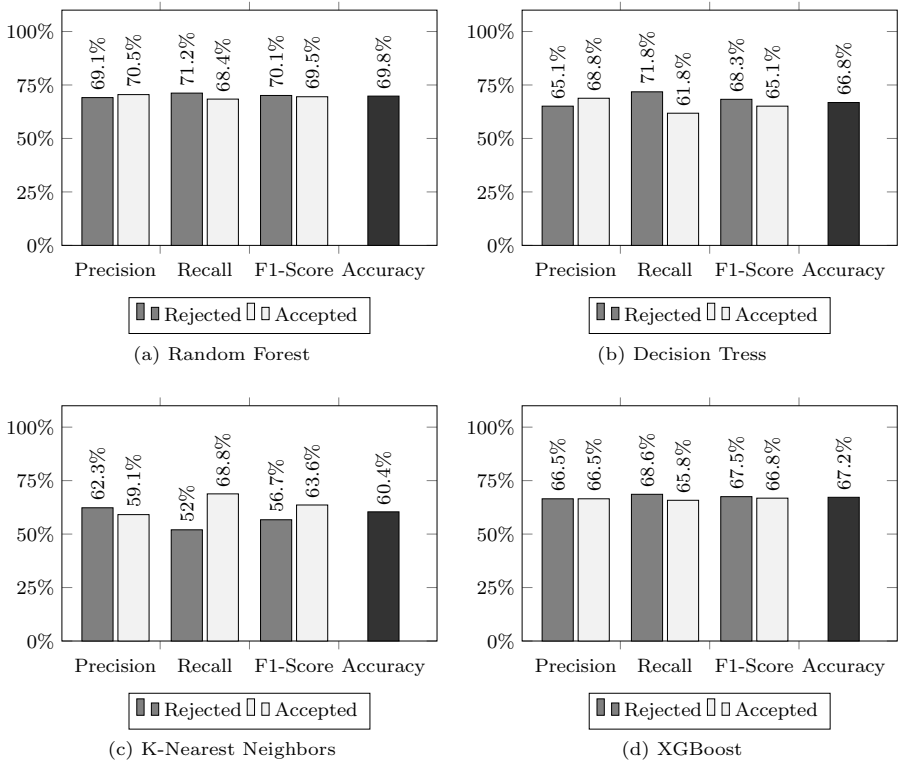
\begin{figure}
	\centering
	
	\subfloat[Random Forest]{
	\resizebox{2.3in}{!}{%
    \begin{tikzpicture}
        \begin{axis}[
    		xtick={1,2,3,4},
    		xticklabels={Precision, Recall, F1-Score, Accuracy
    		},
    		enlarge y limits=false,
    		enlarge x limits=0.20,
    		ymin=0,ymax=110,
    		ybar,
    		ytick={0,25,...,100},
    		yticklabels={0\%,25\%,50\%,75\%,100\%},
    		bar width=0.45cm,
    		width=7cm,
    		height = 5cm,
    		yticklabel style={font=\small},
    		xticklabel style={font=\small, /pgf/number format/fixed},
    		major x tick style = {opacity=0},
    		minor x tick num = 1,    
    		minor tick length=1ex,
    		legend style={
    		    at={(0.5,-0.22)},
    			anchor=north,legend columns=-1
            },
    		nodes near coords style={rotate=90,  anchor=west, font=\small},
            nodes near coords =\pgfmathprintnumber{\pgfplotspointmeta}\%
            ]
            \addplot[plot 0,bar group size={0}{2}]
            coordinates {(1,69.1) (2,71.2) (3,70.1)};
            
            \addplot[plot 1, bar group size={1}{2}]
            coordinates {(1,70.5) (2,68.4) (3,69.5)};
    
            \addplot[plot 2,bar group size={0}{1}]
            coordinates { (4,69.8) };
    
           \legend	{Rejected,
    			Accepted
    		}
        \end{axis}
    \end{tikzpicture}
    \label{fig:ml-model-random-forest}
    }
    }
    \subfloat[Decision Tress]{
	\resizebox{2.3in}{!}{%
    \begin{tikzpicture}
        \begin{axis}[
    		xtick={1,2,3,4},
    		xticklabels={Precision, Recall, F1-Score, Accuracy
    		},
    		enlarge y limits=false,
    		enlarge x limits=0.20,
    		ymin=0,ymax=110,
    		ybar,
    		ytick={0,25,...,100},
    		yticklabels={0\%,25\%,50\%,75\%,100\%},
    		bar width=0.45cm,
    		width=7cm,
    		height = 5cm,
    		yticklabel style={font=\small},
    		xticklabel style={font=\small, /pgf/number format/fixed},
    		major x tick style = {opacity=0},
    		minor x tick num = 1,    
    		minor tick length=1ex,
    		legend style={
    		    at={(0.5,-0.22)},
    			anchor=north,legend columns=-1
            },
    		nodes near coords style={rotate=90,  anchor=west, font=\small},
            nodes near coords =\pgfmathprintnumber{\pgfplotspointmeta}\%
            ]
            \addplot[plot 0,bar group size={0}{2}]
            coordinates {(1,65.1) (2,71.8) (3,68.3)};
            
            \addplot[plot 1, bar group size={1}{2}]
            coordinates {(1,68.8) (2,61.8) (3,65.1)};
    
            \addplot[plot 2,bar group size={0}{1}]
            coordinates { (4,66.8) };
    
           \legend	{Rejected,
    			Accepted
    		}
        \end{axis}
    \end{tikzpicture}
    \label{fig:ml-model-decision-trees}
    }
    }
    
    \subfloat[K-Nearest Neighbors]{
	\resizebox{2.3in}{!}{%
    \begin{tikzpicture}
        \begin{axis}[
    		xtick={1,2,3,4},
    		xticklabels={Precision, Recall, F1-Score, Accuracy
    		},
    		enlarge y limits=false,
    		enlarge x limits=0.20,
    		ymin=0,ymax=110,
    		ybar,
    		ytick={0,25,...,100},
    		yticklabels={0\%,25\%,50\%,75\%,100\%},
    		bar width=0.45cm,
    		width=7cm,
    		height = 5cm,
    		yticklabel style={font=\small},
    		xticklabel style={font=\small, /pgf/number format/fixed},
    		major x tick style = {opacity=0},
    		minor x tick num = 1,    
    		minor tick length=1ex,
    		legend style={
    		    at={(0.5,-0.22)},
    			anchor=north,legend columns=-1
            },
    		nodes near coords style={rotate=90,  anchor=west, font=\small},
            nodes near coords =\pgfmathprintnumber{\pgfplotspointmeta}\%
            ]
            \addplot[plot 0,bar group size={0}{2}]
            coordinates {(1,62.3) (2,52) (3,56.7)};
            
            \addplot[plot 1, bar group size={1}{2}]
            coordinates {(1,59.1) (2,68.8) (3,63.6)};
    
            \addplot[plot 2,bar group size={0}{1}]
            coordinates { (4,60.4) };
    
           \legend	{Rejected,
    			Accepted
    		}
        \end{axis}
    \end{tikzpicture}
    \label{fig:ml-model-knn}
    }
    }
    \subfloat[XGBoost]{
	\resizebox{2.3in}{!}{%
    \begin{tikzpicture}
        \begin{axis}[
    		xtick={1,2,3,4},
    		xticklabels={Precision, Recall, F1-Score, Accuracy
    		},
    		enlarge y limits=false,
    		enlarge x limits=0.20,
    		ymin=0,ymax=110,
    		ybar,
    		ytick={0,25,...,100},
    		yticklabels={0\%,25\%,50\%,75\%,100\%},
    		bar width=0.45cm,
    		width=7cm,
    		height = 5cm,
    		yticklabel style={font=\small},
    		xticklabel style={font=\small, /pgf/number format/fixed},
    		major x tick style = {opacity=0},
    		minor x tick num = 1,    
    		minor tick length=1ex,
    		legend style={
    		    at={(0.5,-0.22)},
    			anchor=north,legend columns=-1
            },
    		nodes near coords style={rotate=90,  anchor=west, font=\small},
            nodes near coords =\pgfmathprintnumber{\pgfplotspointmeta}\%
            ]
            \addplot[plot 0,bar group size={0}{2}]
            coordinates {(1,66.5) (2,68.6) (3,67.5)};
            
            \addplot[plot 1, bar group size={1}{2}]
            coordinates {(1,66.5) (2,65.8) (3,66.8)};
    
            \addplot[plot 2,bar group size={0}{1}]
            coordinates {(4,67.2)};
    
           \legend{Rejected,
    			Accepted
    		}
        \end{axis}
    \end{tikzpicture}
    \label{fig:ml-model-xgboost}
    }
    }

\caption{Performance of our machine learning models.}
\label{fig:performance-machine-models}
\end{figure}

\subsubsection{Model Performance Evaluation}
\label{subsec:performanceEvaluation}

We experiment with our models to see how well the classification models perform based on our features. Fig. \ref{fig:performance-machine-models} summarizes the performance of our models. Our primary focus is to predict the rejected edits. We see that our models can predict the rejected edits with 62.3\%--69.1\% precision. The random forest can predict the rejected edits more precisely than the other three models. Its precision is  69.1\%.
On the other hand, XGBoost shows the highest recall (i.e., 71.8\%). However, the highest F1-score is achieved by the random forest model in predicting rejected edits. The precision to predict the accepted edits ranges from 59.1\%--70.5\%. Like rejected edits, the random forest model achieves the highest precision in predicting accepted edits. On the contrary, the k-nearest neighbors shows the lowest precision.

From Fig. \ref{fig:performance-machine-models}, we see that the overall accuracy of the models is more than 60\%. The highest accuracy is about 70\%. Our experimental result shows that the random forest performs best, whereas the k-nearest neighbors shows the lowest performance. The k-nearest neighbors could suffer from the high dimensional data. However, according to the experimental result, XGBoost slightly outperforms decision trees but underperforms the random forest. We thus further investigate why XGBoost does not outperform random forest. We analyze the predicted class of XGBoost and random forest models against our test dataset. We find that XGBoost misclassified 31 samples (15 accepted + 16 rejected), which random forests classified correctly.

In our dataset, addition/removal of duplication notes get rejected more than 94\% times, and signatures get rejected 100\% times. However, a few samples are classified as accepted by XGBoost even after the addition/removal of duplication notes or signatures. On the contrary, random forest classified them correctly as rejected. Besides, XGBoost classified several samples with trivial text changes as rejected, which were incorrect. Our analysis shows that trivial changes have more chance of being accepted than rejected. However, random forest classified the target class correctly in those cases. Hence, the above reasons could explain why XGBoost slightly underperforms random forest. Therefore, we select the random forest model to deploy with our online tool.



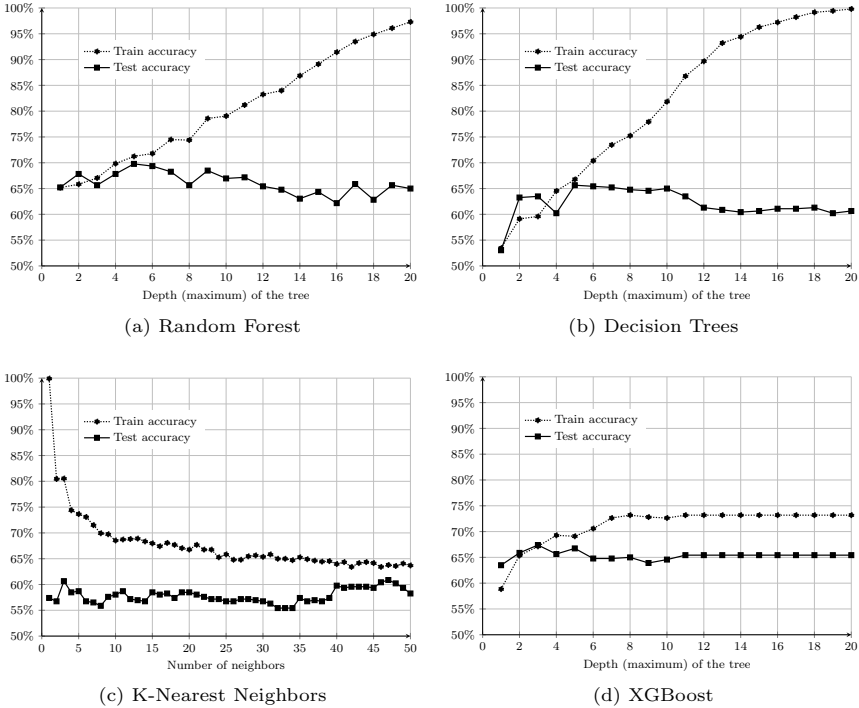
\begin{figure}[!htb]
\centering

    \subfloat[Random Forest]{
	\resizebox{2.2in}{!}{%
    \begin{tikzpicture}
    \begin{axis}[
        width=4in,
        height=3in,
        axis lines=left,
        grid=both,
        legend style={draw=none, font=\small},
        legend cell align=left,
        legend style={
        	at={(0.10,0.80)},
        	font=\small,
            anchor=west,
        },
        label style={font=\small},
        ticklabel style={font=\small},
        xmin=0,
        ymin=50,
        ymax=100,
        xmax=20,
        ytick={0,5,10,...,100},
    	yticklabels={0\%,5\%,10\%,15\%,20\%,25\%,30\%,35\%,40\%,45\%,50\%,55\%,60\%,65\%,70\%,75\%,80\%,85\%,90\%,95\%,100\%},
        xlabel=Depth (maximum) of the tree,
        legend entries={Train accuracy, 
        Test accuracy,
        },
    ]
    \addplot[color=black, mark=*, thick, mark options={scale=0.8},densely dotted] 
    coordinates {(1, 65.18)
                (2, 65.83)
                (3, 67.04)
                (4, 69.83)
                (5, 71.23)
                (6, 71.79)
                (7, 74.49)
                (8, 74.39)
                (9, 78.58)
                (10, 79.05)
                (11, 81.19)
                (12, 83.24)
                (13, 83.99)
                (14, 86.87)
                (15, 89.11)
                (16, 91.43)
                (17, 93.48)
                (18, 94.88)
                (19, 96.09)
                (20, 97.30)};
    
    \addplot[color=black, mark=square*, thick, mark options={scale=0.8}]
    coordinates {(1, 65.22)
                (2, 67.83)
                (3, 65.65)
                (4, 67.83)
                (5, 69.78)
                (6, 69.35)
                (7, 68.26)
                (8, 65.65)
                (9, 68.48)
                (10, 66.96)
                (11, 67.17)
                (12, 65.43)
                (13, 64.78)
                (14, 63.04)
                (15, 64.35)
                (16, 62.17)
                (17, 65.87)
                (18, 62.83)
                (19, 65.65)
                (20, 65.00)};

    \end{axis}
    \end{tikzpicture}
    \label{fig:tuning-random-forest}
    }
    }
    \subfloat[Decision Trees]{
	\resizebox{2.2in}{!}{%
    \begin{tikzpicture}
    \begin{axis}[
        width=4in,
        height=3in,
        axis lines=left,
        grid=both,
        legend style={draw=none, font=\small},
        legend cell align=left,
        legend style={
        	at={(0.10,0.80)},
        	font=\small,
            anchor=west,
        },
        label style={font=\small},
        ticklabel style={font=\small},
        xmin=0,
        ymin=50,
        ymax=100,
        xmax=20,
        ytick={0,5,10,...,100},
    	yticklabels={0\%,5\%,10\%,15\%,20\%,25\%,30\%,35\%,40\%,45\%,50\%,55\%,60\%,65\%,70\%,75\%,80\%,85\%,90\%,95\%,100\%},
        xlabel=Depth (maximum) of the tree,
        legend entries={Train accuracy, 
        Test accuracy,
        },
    ]
    \addplot[color=black, mark=*, thick, mark options={scale=0.8},densely dotted]
    coordinates {(1, 53.45)
                (2, 59.12)
                (3, 59.59)
                (4, 64.53)
                (5, 66.76)
                (6, 70.39)
                (7, 73.46)
                (8, 75.23)
                (9, 77.93)
                (10, 81.84)
                (11, 86.78)
                (12, 89.66)
                (13, 93.20)
                (14, 94.41)
                (15, 96.28)
                (16, 97.21)
                (17, 98.23)
                (18, 99.16)
                (19, 99.44)
                (20, 99.81)};
    
    \addplot[color=black, mark=square*, thick, mark options={scale=0.8}]
    coordinates {(1, 53.04)
                (2, 63.26)
                (3, 63.48)
                (4, 60.22)
                (5, 65.65)
                (6, 65.43)
                (7, 65.22)
                (8, 64.78)
                (9, 64.57)
                (10, 65.00)
                (11, 63.48)
                (12, 61.30)
                (13, 60.87)
                (14, 60.43)
                (15, 60.65)
                (16, 61.09)
                (17, 61.09)
                (18, 61.30)
                (19, 60.22)
                (20, 60.65)};

    \end{axis}
    \end{tikzpicture}
    \label{fig:tuning-decision-trees}
    }
    }
    
    \subfloat[K-Nearest Neighbors]{
	\resizebox{2.2in}{!}{%
    \begin{tikzpicture}
    \begin{axis}[
        width=4in,
        height=3in,
        axis lines=left,
        grid=both,
        legend style={draw=none, font=\small},
        legend cell align=left,
        legend style={
        	at={(0.10,0.80)},
        	font=\small,
            anchor=west,
        },
        label style={font=\small},
        ticklabel style={font=\small},
        xmin=0,
        xmax=50,
        ymin=50,
        ymax=100,
        xtick={0,5,10,15,20,25,30,35,40,45,50},
        ytick={0,5,10,...,100},
    	yticklabels={0\%,5\%,10\%,15\%,20\%,25\%,30\%,35\%,40\%,45\%,50\%,55\%,60\%,65\%,70\%,75\%,80\%,85\%,90\%,95\%,100\%},
        xlabel=Number of neighbors,
        legend entries={Train accuracy, 
        Test accuracy,
        },
    ]
    \addplot[color=black, mark=*, thick, mark options={scale=0.8},densely dotted]
    coordinates {(1, 99.91)
                (2, 80.45)
                (3, 80.54)
                (4, 74.39)
                (5, 73.65)
                (6, 73.09)
                (7, 71.51)
                (8, 69.93)
                (9, 69.74)
                (10, 68.53)
                (11, 68.72)
                (12, 68.81)
                (13, 68.90)
                (14, 68.34)
                (15, 67.97)
                (16, 67.41)
                (17, 68.06)
                (18, 67.69)
                (19, 67.04)
                (20, 66.76)
                (21, 67.69)
                (22, 66.76)
                (23, 66.76)
                (24, 65.27)
                (25, 65.83)
                (26, 64.80)
                (27, 64.80)
                (28, 65.46)
                (29, 65.64)
                (30, 65.36)
                (31, 65.83)
                (32, 64.99)
                (33, 64.99)
                (34, 64.71)
                (35, 65.27)
                (36, 64.90)
                (37, 64.62)
                (38, 64.43)
                (39, 64.53)
                (40, 63.97)
                (41, 64.34)
                (42, 63.41)
                (43, 64.15)
                (44, 64.34)
                (45, 64.15)
                (46, 63.41)
                (47, 63.78)
                (48, 63.59)
                (49, 64.06)
                (50, 63.69)};
                    
    \addplot[color=black, mark=square*, thick, mark options={scale=0.8}]
    coordinates {(1, 57.39)
                (2, 56.74)
                (3, 60.65)
                (4, 58.48)
                (5, 58.70)
                (6, 56.74)
                (7, 56.52)
                (8, 55.87)
                (9, 57.61)
                (10, 58.04)
                (11, 58.70)
                (12, 57.17)
                (13, 56.96)
                (14, 56.74)
                (15, 58.48)
                (16, 58.04)
                (17, 58.26)
                (18, 57.39)
                (19, 58.48)
                (20, 58.48)
                (21, 58.04)
                (22, 57.61)
                (23, 57.17)
                (24, 57.17)
                (25, 56.74)
                (26, 56.74)
                (27, 57.17)
                (28, 57.17)
                (29, 56.96)
                (30, 56.74)
                (31, 56.30)
                (32, 55.43)
                (33, 55.43)
                (34, 55.43)
                (35, 57.39)
                (36, 56.74)
                (37, 56.96)
                (38, 56.74)
                (39, 57.39)
                (40, 59.78)
                (41, 59.35)
                (42, 59.57)
                (43, 59.57)
                (44, 59.57)
                (45, 59.35)
                (46, 60.43)
                (47, 60.87)
                (48, 60.22)
                (49, 59.35)
                (50, 58.26)};

    \end{axis}
    \end{tikzpicture}
    \label{fig:tuning-knn}
    }
    }
    \subfloat[XGBoost]{
	\resizebox{2.2in}{!}{%
    \begin{tikzpicture}
    \begin{axis}[
        width=4in,
        height=3in,
        axis lines=left,
        grid=both,
        legend style={draw=none, font=\small},
        legend cell align=left,
        legend style={
        	at={(0.10,0.80)},
        	font=\small,
            anchor=west,
        },
        label style={font=\small},
        ticklabel style={font=\small},
        xmin=0,
        ymin=50,
        ymax=100,
        xmax=20,
        ytick={0,5,10,...,100},
    	yticklabels={0\%,5\%,10\%,15\%,20\%,25\%,30\%,35\%,40\%,45\%,50\%,55\%,60\%,65\%,70\%,75\%,80\%,85\%,90\%,95\%,100\%},
        xlabel=Depth (maximum) of the tree,
        legend entries={Train accuracy, 
        Test accuracy,
        },
    ]
   \addplot[color=black, mark=*, thick, mark options={scale=0.8},densely dotted]
    coordinates {(1, 58.85)
                (2, 65.36)
                (3, 67.13)
                (4, 69.27)
                (5, 69.09)
                (6, 70.58)
                (7, 72.63)
                (8, 73.18)
                (9, 72.81)
                (10, 72.63)
                (11, 73.18)
                (12, 73.18)
                (13, 73.18)
                (14, 73.18)
                (15, 73.18)
                (16, 73.18)
                (17, 73.18)
                (18, 73.18)
                (19, 73.18)
                (20, 73.18)};
    
    \addplot[color=black, mark=square*, thick, mark options={scale=0.8}]
    coordinates {(1, 63.48)
                (2, 65.87)
                (3, 67.39)
                (4, 65.65)
                (5, 66.74)
                (6, 64.78)
                (7, 64.78)
                (8, 65.00)
                (9, 63.91)
                (10, 64.57)
                (11, 65.43)
                (12, 65.43)
                (13, 65.43)
                (14, 65.43)
                (15, 65.43)
                (16, 65.43)
                (17, 65.43)
                (18, 65.43)
                (19, 65.43)
                (20, 65.43)};

    \end{axis}
    \end{tikzpicture}
    \label{fig:tuning-xgboost}
    }
    }
    
\caption{Parameter tuning to reduce model overfitting.}
\label{fig:tuning-ml-models}
\end{figure}

Our models could capture unnecessary details or too specific relationships within the training dataset and thus suffer from overfitting. However, overfit models are not very stable since they fail to generalize well to the data. Moreover, overfit models generally perform poorly on unseen (e.g., test) data. In particular, a overfit model has a substantial difference between the accuracy of the training and test dataset. That is why we can never trust an overfit model and put it into deployment. We thus attempt to reduce overfitting before deploying our model. In particular, we tuned the critical parameters of the model to balance the accuracy between training and test datasets. Then, we set the values of the parameters to avoid model overfitting.

\begin{table}[htb]
\centering
	\caption{Accuracy of the machine learning models}
	\label{table:ml-model-accuracy}
 	\resizebox{2.5in}{!}{%
    \begin{tabular}{l|l|l}
    \toprule
    \multicolumn{1}{c|}{\multirow{2}{*}{\textbf{Model Name}}} & \multicolumn{2}{c}{\textbf{Accuracy}}                                         \\ \cmidrule{2-3} 
    \multicolumn{1}{c|}{}                                                     & \multicolumn{1}{c|}{\textbf{Testing}} & \multicolumn{1}{c}{\textbf{Training}} \\ \toprule
    \textbf{Random Forest}                                                     &  69.8\%  & 71.2\%  \\ \midrule
    \textbf{Decision Trees}                                                                 &  66.8\%  & 66.8\%  \\ \midrule
    \textbf{K-Nearest Neighbors}                                                                 &  60.4\%  & --       \\ \midrule
    \textbf{XGBoost}                                                                 &  67.2\%  & 67.1\%  \\ \bottomrule
    \end{tabular}
    }
\end{table}

Fig. \ref{fig:tuning-ml-models} shows the model accuracies on the training and test dataset in contrast to example critical parameters of the model. For example, we run the random forest model over the depth of the tree from $1$ to $20$. As shown in Fig. \ref{fig:tuning-random-forest}, the training accuracy improves with the depth of the tree. However, the difference between training and test accuracies increases when the depth exceeds five. We thus set the depth value as five while training the random forest model. We set the depth values for the decision trees and XGBoost models similarly. Their depth values are five and three. In particular, we attempt to balance the training and test accuracies when determining depth values.

On the other hand, Fig. \ref{fig:tuning-knn} shows the training and test accuracies of the k-nearest neighbors in contrast to the number of neighbors. We vary the number of neighbors from $1$ to $50$, evaluate the model on the train and test datasets for each number of neighbors, and report the accuracy. We see that performance on the test set improves initially and then worsens, and performance on the training set continues to degrade. As shown in Fig. \ref{fig:tuning-knn}, the training accuracy is dropping to converge with the line for the test set. However, the training and testing accuracies are very close when the number of neighbors is $46$. We thus choose the parameter value is $46$ (i.e., number of neighbors $= 46$).


    
    

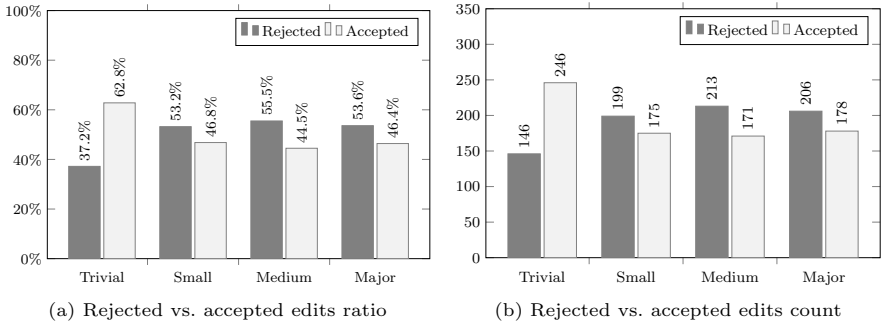
\begin{figure}[!htb]
\centering
   	\pgfplotstableread{
    		1	37.2   62.8   146  246  
    		2	53.2   46.8   199  175
    		3	55.5   44.5    213  171
    		4	53.6   46.4    206  178

    	}\datatable
      \subfloat[Rejected vs. accepted edits ratio]{
      \label{fig:rejected-vs-accepted-edit-ratio}
      \resizebox{2.25in}{!}{%
      \begin{tikzpicture}
        	\begin{axis}[
        	xtick=data,
        	xticklabels={Trivial, Small, Medium, Major},
        	enlarge y limits=false,
        	enlarge x limits=0.2,
        	ymin=0,ymax=100,
        	ybar,
        	bar width=0.6cm,
        	width=3.5in,
        	height = 2.5in,
        	ytick={0,20,...,100},
            yticklabels={0\%,20\%,40\%,60\%,80\%,100\%,},
        	ymajorgrids=false,
        	major x tick style = {opacity=0},
        	minor x tick num = 1,    
        	minor tick length=1ex,
        	legend style={
         	legend pos=north east,
        	legend cell align=left,
        	legend columns=-1
            },
            nodes near coords style={rotate=90,  anchor=west}, 
        	nodes near coords =\pgfmathprintnumber{\pgfplotspointmeta}\%
        	]
        	\addplot[draw=black!50, fill=black!50] table[x index=0,y index=1] \datatable;
        	\addplot[draw=black!50, fill=black!5] table[x index=0,y index=2] \datatable;

            \legend	{Rejected,
            		 Accepted
            		 }
        	\end{axis}
    	\end{tikzpicture}
     	}
     	}
      \subfloat[Rejected vs. accepted edits count]{
      \label{fig:rejected-vs-accepted-edit-count}
      \resizebox{2.25in}{!}{%
      \begin{tikzpicture}
        	\begin{axis}[
        	xtick=data,
        	xticklabels={Trivial, Small, Medium, Major},
        	enlarge y limits=false,
        	enlarge x limits=0.2,
        	ymin=0,ymax=350,
        	ybar,
        	bar width=0.6cm,
        	width=3.5in,
        	height = 2.45in,
        	ytick={0,50,...,350},
        	ymajorgrids=false,
        	major x tick style = {opacity=0},
        	minor x tick num = 1,    
        	minor tick length=1ex,
        	legend style={
         	legend pos=north east,
        	legend cell align=left,
        	legend columns=-1
            },
            nodes near coords style={rotate=90,  anchor=west}, 
        	nodes near coords =\pgfmathprintnumber{\pgfplotspointmeta}
        	]
        	\addplot[draw=black!50, fill=black!50] table[x index=0,y index=3] \datatable;
        	\addplot[draw=black!50, fill=black!5] table[x index=0,y index=4] \datatable;

            \legend	{Rejected,
            		 Accepted
            		 }
        	\end{axis}
    	\end{tikzpicture}
     	}
     	}
\caption{Change of post contents by edits.}
\label{fig:change-of-posts}
\end{figure}

\begin{figure}
	\centering
	\subfloat[Classifier that rejects trivial edits]{
    \label{fig:rejects-trivial-edits}
    \resizebox{2.25in}{!}{%
    \begin{tikzpicture}
        \begin{axis}[
    		xtick={1,2,3,4},
    		xticklabels={Precision, Recall, F1-Score, Accuracy
    		},
    		enlarge y limits=false,
    		enlarge x limits=0.20,
    		ymin=0,ymax=120,
    		ybar,
    		ytick={0,25,...,100},
    		yticklabels={0\%,25\%,50\%,75\%,100\%},
    		bar width=0.6cm,
    		width=3.5in,
    		height = 2.5in,
    		yticklabel style={font=\small},
    		xticklabel style={font=\small, /pgf/number format/fixed},
    		major x tick style = {opacity=0},
    		minor x tick num = 1,    
    		minor tick length=1ex,
    		legend style={
         	legend pos=north east,
        	legend cell align=left,
        	legend columns=-1
            },
    		nodes near coords style={rotate=90,  anchor=west, font=\small},
            nodes near coords =\pgfmathprintnumber{\pgfplotspointmeta}\%
            ]
            \addplot[plot 0,bar group size={0}{2}]
            coordinates {(1,37.2) (2,19.1) (3,25.2)};
            
            \addplot[plot 1, bar group size={1}{2}]
            coordinates {(1,45.9) (2,68.1) (3,54.8)};
    
            \addplot[plot 2,bar group size={0}{1}]
            coordinates {(4,43.7)};
    
           \legend{Rejected,
    			Accepted
    		}
        \end{axis}
    \end{tikzpicture}
    }
    }
    \subfloat[Classifier that rejects non-trivial edits]{
    \label{fig:accepts-trivial-edits}
    \resizebox{2.25in}{!}{%
    \begin{tikzpicture}
        \begin{axis}[
    		xtick={1,2,3,4},
    		xticklabels={Precision, Recall, F1-Score, Accuracy
    		},
    		enlarge y limits=false,
    		enlarge x limits=0.20,
    		ymin=0,ymax=120,
    		ybar,
    		ytick={0,25,...,100},
    		yticklabels={0\%,25\%,50\%,75\%,100\%},
    		bar width=0.6cm,
    		width=3.5in,
    		height = 2.5in,
    		yticklabel style={font=\small},
    		xticklabel style={font=\small, /pgf/number format/fixed},
    		major x tick style = {opacity=0},
    		minor x tick num = 1,    
    		minor tick length=1ex,
    		legend style={
         	legend pos=north east,
        	legend cell align=left,
        	legend columns=-1
            },
    		nodes near coords style={rotate=90,  anchor=west, font=\small},
            nodes near coords =\pgfmathprintnumber{\pgfplotspointmeta}\%
            ]
            \addplot[plot 0,bar group size={0}{2}]
            coordinates {(1,54.1) (2,80.9) (3,64.9)};
            
            \addplot[plot 1, bar group size={1}{2}]
            coordinates {(1,62.8) (2,32.0) (3,42.3)};
    
            \addplot[plot 2,bar group size={0}{1}]
            coordinates {(4,56.3)};
    
           \legend{Rejected,
    			Accepted
    		}
        \end{axis}
    \end{tikzpicture}
    }
    }
\caption{Performance of baseline machine learning model.}
\label{fig:performance-baseline-model}
\end{figure}
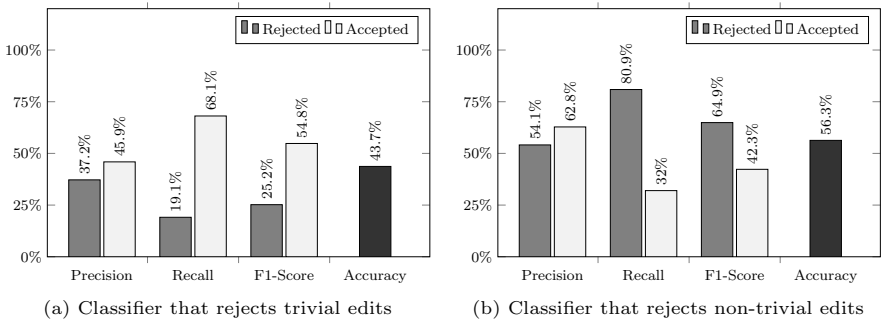

\subsubsection{Performance of Baseline Model}
\label{subsec:baselineModel}

To the best of our knowledge, there was no existing machine learning model to identify the rejected edits with reasons at the time when we conducted this study. However, we construct baseline models that reject/accept trivial edits and evaluate their performance. We thus first attempt to categorize the types of edits conducted in the samples of our dataset. We calculate the \emph{Levenshtein} distance between the original post's content (text + code) and the post with suggested edits. We normalize the distance with respect to the character length of the original posts. Then, we classify the edits into four categories based on the edit distance. They are -- \emph{trivial} (distance $\leq$ lower quartile), \emph{small} (lower quartile $>$ distance $\leq$ median), \emph{medium} (median $>$ distance $\leq$ upper quartile) and \emph{major} (distance $>$ upper quartile) edit.

Fig. \ref{fig:change-of-posts} shows the percentage and count of rejected and accepted edits for each category. We see that 62.8\% of trivial edits get accepted (Fig. \ref{fig:rejected-vs-accepted-edit-ratio}) in our dataset. On the contrary, such a statistic is only 37.2\% for rejected edits. However, the percentage of rejected edits is higher than the accepted edits for the remaining three categories. We then develop a rule-based classifier that rejects trivial edits. Fig. \ref{fig:rejects-trivial-edits} summarizes the performance of the classifier. The precision of identifying rejected edits is below 40\%. The classifier also has poor recall (i.e., 19.1\%) in identifying rejected edits with 43.7\% overall accuracy. Next, we build a classifier that rejects non-trivial (i.e., small, medium \& major) edits. That is, it accepts trivial edits. Fig. \ref{fig:accepts-trivial-edits} shows its performance. This classifier shows a higher recall in identifying rejected edits than our proposed models. However, the precision, F1-score, and overall accuracy are significantly lower than our best-performing model. Such performances suggest that rejected edits cannot be identified reasonably well only based-on edit categories.

\begin{figure}[!htp]
	\centering
    \resizebox{3.5in}{!}{
    \begin{tikzpicture}
    \begin{axis}[
        xmin=0,
        width=4in,
        height=3.2in,
        enlarge y limits=0.14,
        bar width=0.45cm,
        symbolic y coords={Gratitude, Greetings, Deprecation, Duplication, Signature, Inactive Hyperlink, Deface Post},
        ytick=data,
        xtick={0,20,...,100},
        xticklabels={0\%,20\%,40\%,60\%,80\%,100\%,},
        xbar stacked,
        legend style={at={(0.5,-0.20)},
          anchor=north,legend columns=-1},
        ytick=data,
        every node near coord/.style={
            font=\small,
            black!100,
        },
        nodes near coords align=center,
        nodes near coords style={yshift= 0.4cm, rotate=0},
        xbar stacked, nodes near coords={\hspace{0cm} \pgfmathprintnumber[precision=1]{\pgfplotspointmeta}\%}
        ]
    \addplot+[xbar, fill=gray!80,draw=gray!80] plot coordinates {(87.5,Gratitude) (69.2,Greetings) (100,Deprecation) (94.1,Duplication) (100,Signature) (76.2,Inactive Hyperlink) (87.5,Deface Post)};
    \addplot+[xbar, fill=gray!20,draw=gray!80] plot coordinates {(12.5,Gratitude) (30.8,Greetings) (0,Deprecation) (5.9,Duplication) (0,Signature) (23.8,Inactive Hyperlink) (12.5,Deface Post)};
    
    \legend{Rejected, Accepted}
    
    \end{axis}
    \end{tikzpicture}
    }
    \caption{Percentage of rejected and accepted edits.}
	\label{fig:rejected-vs-accepted-ratio}
\end{figure}
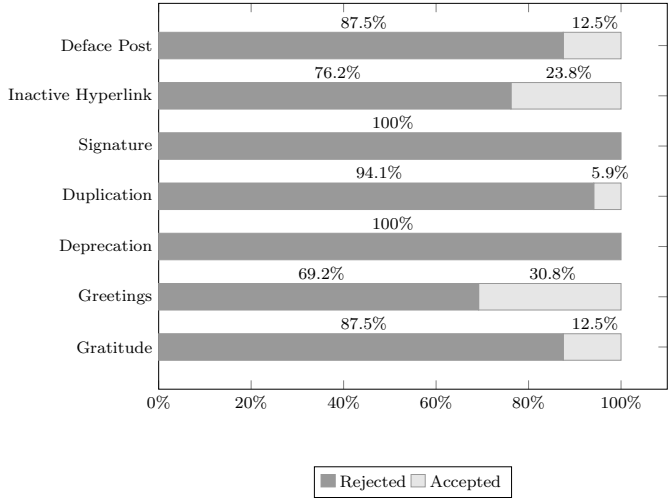

\subsection{Rejection Reason Classifier}
\label{subsec:rejectionReasonClassifier}

In the previous section, we evaluate the performance of the machine learning models in predicting the rejected edits. This section analyzes how accurately the rejection reason classifier can identify the potential reasons for those rejections.

Our rejected reason classifier can almost accurately identify several rejection reasons by applying the same approach as we extracted features \ref{subsec:featureExtraction}. For example, our manual investigation identified nine keywords (e.g., thanks, welcome) (Section \ref{subsec:featureExtraction}) that were utilized to identify gratitude. However, our further analysis finds that the addition or removal of gratitude is rejected 85.5\% times (e.g., Fig. \ref{fig:rejected-vs-accepted-ratio}) in our dataset. We thus used this lightweight keyword-based technique to identify the reason ``gratitude add/remove'' when the edits are predicted as rejected. However, to avoid multiple computations, we look back at the feature vector to check the feature value of gratitude. 
Similarly, we identify the following potential edits rejection reasons by analyzing the feature vector.

\begin{tcolorbox}[colframe=black!50, colback=white,left=1pt,right=1pt,top=1pt,bottom=1pt, arc=1pt]

    Undesired text/code formatting, complete change of post, deface post, gratitude add/remove, greetings add/remove, signature add/remove, deprecation note add/remove, duplication note add/remove, undesired reference modification

\end{tcolorbox}

\begin{table}[!t]
	\centering
	\captionsetup{justification=centering, labelsep=newline}
	\caption{Performance of classifiers to identify undesired text/code addition/deletion}
	\label{table:undesired-text-code-add-delete-reasons}
	\resizebox{4in}{!}{%
    \begin{tabular}{l|c|c|c|c}
    \toprule
    \textbf{Classifier} & \textbf{Precision} & \textbf{Recall} & \textbf{F1-Score} & \textbf{Accuracy} \\ \midrule
    \textbf{Undesired Text Addition}    & 69.2\% & 48.1\% & 56.7\% & 63.4\%
    \\ \midrule
    \textbf{Undesired Text Removal}    & 62.1\% & 50.8\% & 55.9\% & 58.0\% \\ \midrule
    \textbf{Undesired Code Addition}    & 65.3\%  & 64.0\% & 64.6\% & 65.0\% \\ \midrule
    \textbf{Undesired Code Removal}    & 69.0\% & 74.4\% & 71.6\% & 68.1\% \\ \bottomrule
    \end{tabular}
    }
\end{table}

For a remaining few rejection reasons, such as undesired text/code addition/removal, we primarily apply n-grams and pos-tagging-based techniques. Unfortunately, we did not find satisfactory performance (e.g., accuracy $<$ 50\%). 
Our further investigation suggests that both desired and undesired texts contain similar words, phrases, or patterns. Therefore, we cannot distinguish them using words, phrases, or pos-based patterns. However, the character length of added or removed text/code can identify the undesired addition or removal of text/code reasonably well. We thus extract the added/removed text or code from our manually analyzed dataset using the appropriate \texttt{HTML} tags. In particular, we extract the contents of \texttt{HTML} elements with a class attribute with either \texttt{diff-add} \& \texttt{diff-delete} values. We measure the length of added/removed characters of text or code. We then normalize the length with respect to the total length of the text/code of revisions. Next, we separate samples into two classes according to our manual label -- (1) undesired vs. desired text addition, (2) undesired vs. desired text removal, (3) undesired vs. desired code addition, and (4) undesired vs. desired code removal. We developed four random forest classifiers to identify the undesired text/code addition/removal. However, we resolve the class imbalance problem using Synthetic Minority Oversampling Technique (SMOTE) \citep{wang2006classification}.
Table \ref{table:undesired-text-code-add-delete-reasons} shows the performance of the classifiers. The precision of identifying the undesired text/code addition/removal is about 62.1\%--69.2\%. The recall is slightly lower for undesired text addition/removal than undesired code addition/removal. However, the overall accuracy is more than 63\%, except for undesired text removal.



We then evaluate the overall performance of the rejection reason classifier in identifying the potential rejection reasons. We experiment with the rejection reason classifier using our test dataset. In particular, we label each sample in a file as follows.

\begin{table}[!htb]
	\centering
	\captionsetup{justification=centering, labelsep=newline}
	\caption{Confusion matrix and performance of our model to identify the rejection reasons \small{(TP: True Posisitve, TN: True Negative, FP: False Positive \& FN: False Negative)}}
	\label{table:confusion-matrix-performance}
	\resizebox{4in}{!}{%
    \begin{tabular}{c|c|c|c|c|c|c|c}
    \toprule
    \multirow{2}{*}{\textbf{TP}} & \multirow{2}{*}{\textbf{FP}} & \multirow{2}{*}{\textbf{TN}} & \multirow{2}{*}{\textbf{FN}} & \multicolumn{4}{c}{\textbf{Performance}}  \\ 
    
     &  &  &  & \textbf{Precision} & \textbf{Recall} & \textbf{F1-Score} & \textbf{Accuracy} \\ \midrule
    
    129 & 78 & 153 & 63 & 62.3\% & 67.2\% & 64.7\% & 66.7\%  \\ \bottomrule
    \end{tabular}
    }
\end{table}

(1) \textbf{Identified:} the reasons detected by the rejection reason classifier.
(2) \textbf{Expected:} the actual reasons based on our manual analysis.
\noindent We then create a confusion matrix to analyze the performance of the rejection reason classifier as follows.
(i) \emph{True Positive (TP)} = `identified' reasons = `expected' reasons
(ii) \emph{False Positive (FP)} =  (`identified' reasons $\neq$ `expected' reasons)  or (`identified' reasons but `expected' no reasons)
(iii) \emph{True Negative (TN)} = `identified' no reasons and `expected' no reasons, and 
(iv) \emph{False Negative (FN)} = `identified' no reasons but `expected' one or more reasons. 
Using the above matrix, we compute four standard metrics 
(\emph{Precision, Recall, F1-score,} and \emph{Accuracy}) to compute the performance of the model to identify the rejected reasons. 

Table \ref{table:confusion-matrix-performance} shows the confusion matrix and performance of rejection reason classifier to identify the potential reasons for edit rejections. Our analysis shows that our rejection reason classifier can identify the potential rejection reasons with 62.3\% precision, 67.2\% recall, 64.7\% F1-score, and 66.7\% overall accuracy. 
We further analyze in which cases our rejection reason classifier fails to identify rejection reasons. We find that our model mainly fails to identify all the reasons when there were multiple reasons to reject suggested edits. In that cases, our model can partly identify the reasons. For example, there are three reasons for rejection. However, our model identifies two of them accurately. 
Note that we identify the potential rejection reason as community mistrust when -- (1) our rejection reason classifier cannot identify any reason and (2) the reputation score of the user who suggested edits is below 2K. We choose a reputation score $< 2K$ since users with a reputation score $< 2K$ have no privilege to edit posts instantly.
However, our rejection reason classifier cannot identify a couple of reasons (e.g., incorrect text/code change). We will discuss them in Section \ref{sec:deviation}.

\smallskip\begin{tcolorbox}[colframe=black!50,colback=white,left=0pt,right=1pt,top=1pt,bottom=1pt, boxrule=1pt,arc=1pt, width=4.7in, center]

    \textbf{Summary of $RQ_1$ and hypothesis testing:} We developed four machine learning classifiers based on 15 texts \& user-based features to predict rejected edits. Our best-performing classifier (i.e., random forest) can predict the rejected edits with 69.1\% precision, 71.2\% recall, 70.1\% F1-score, and 69.8\% overall accuracy. We then develop a classifier to identify the potential reasons behind rejections. Our rejected reason classifier can identify the potential rejection reasons with overall 66.7\% accuracy. 
    Such findings \emph{reject} our null hypothesis (i.e., \emph{$H_1$}), which hypothesized that the accuracy of our developed classifiers is not better than 50\%. 
    
\end{tcolorbox}


\section{EditEx: A Recommender for Early Fixes to Suggested Edits}
\label{sec:editex-tool}

We can assess the actual impact of our developed classifier if it can automatically assist users during their editing of SO posts. Editing is a time-consuming and largely voluntary activity in SO. Therefore, users can be assisted with a tool that can recommend potential fixes to their undesired edits. We thus focus on introducing an online tool, namely \texttt{EditEx}, that interacts with our classifier, identifies the potential rejection reasons from the suggested edits, and helps to reduce the likelihood of rejection of suggested edits. We then assess the tool's effectiveness via real-world usage by users and answer the research question as follows.

\begin{tcolorbox}[colframe=black!70,colback=white,left=0pt,right=1pt,top=1pt,bottom=1pt, boxrule=1pt,arc=1pt, width=4.7in, center]
    \textbf{$RQ_2$)} To what extent can \texttt{EditEx} be helpful to users avoid rejection in their edits?
\end{tcolorbox}

For $RQ_2$, we define the following null hypothesis.

\begin{tcolorbox}[colframe=black!70,colback=white,left=0pt,right=1pt,top=1pt,bottom=1pt, boxrule=1pt,arc=1pt, width=4.7in, center]
    \textbf{$H_2$:} Our \texttt{EditEx} tool is not effective in helping users avoid rejected edits.
\end{tcolorbox}

\subsection{\texttt{EditEx} Architecture}
\label{subsec:editex-architecture}

\begin{figure}[htb]
	\centering
	\includegraphics[width=4in]{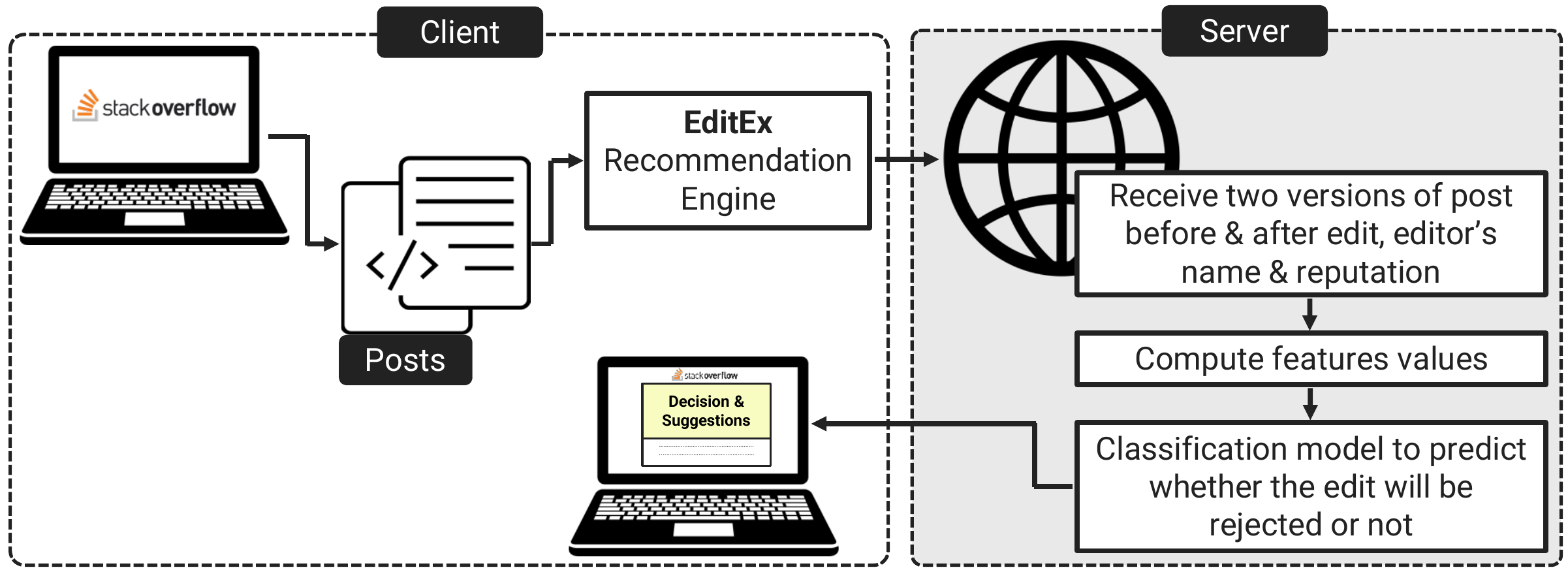}
	\caption{An overview of the \texttt{EditEx} system architecture.}
	\label{fig:editex-system-architecture}
\end{figure}

Fig. \ref{fig:editex-system-architecture} shows an overview of the \texttt{EditEx} architecture. \texttt{EditEx} has two parts: \emph{client} and \emph{server}. On the client-side, users get the \texttt{EditEx} interface. \texttt{EditEx} interface comprises two buttons: \emph{EditEx} and \emph{Suggestion}. EditEx enables users to edit the SO posts. On the other hand, users can check whether the edits will be rejected/accepted and the potential rejection reasons upon rejection by clicking the Suggestion button.

When users click the \emph{Suggestion} button, the client-side script captures the necessary data required to extract the features for the machine learning model. In particular, it captures -- (1) text before edit, (2) text after edit, (3) reputation, and (4) the name of the user who suggests edits. Then it sends this data to the server-side application. Client-side script is written by JavaScript, and server-side application is developed by Java. However, the server-side application captures all the feature values (Section \ref{subsec:featureExtraction}) using the data sent by the client. The feature values form a feature vector. The feature vector is then passed to the classification model to predict whether the edit will be rejected or not. 

The potential rejection reasons are identified upon rejection based on the feature vector and texts (before and after edit). Then, the decision of the edit (i.e., rejected/accepted) and the helpful suggestions that notify the rejection reasons (if rejected) are sent to the client-side script. Finally, the client-side script offers users the result and suggestions.



\begin{figure}[h]
	\centering
	\includegraphics[width=4.7in]{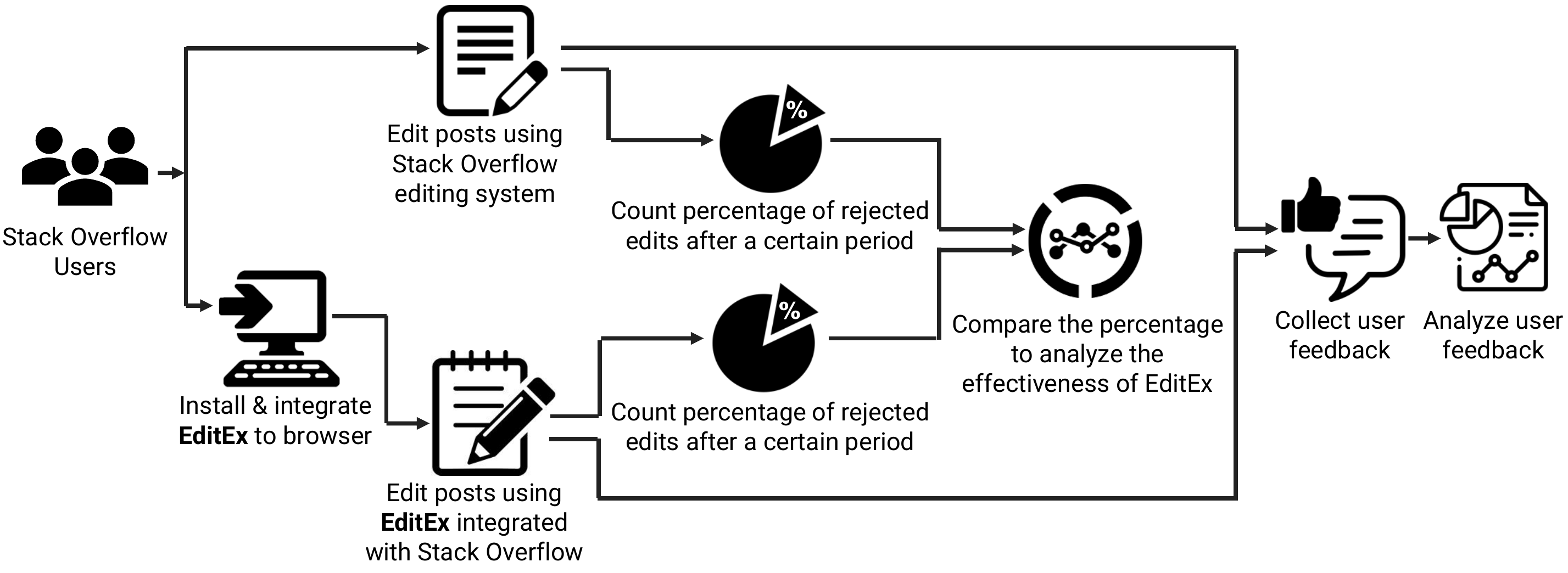}
	\caption{An overview of the \texttt{EditEx} and its effectiveness evaluation.}
	\label{fig:editex-assessment-methodology}
\end{figure}

\subsection{Effectiveness Evaluation Plan of \texttt{EditEx}}
\label{subsec:editex-effectiveness-evaliation-plan}

Fig. \ref{fig:editex-assessment-methodology} shows the overview of how \texttt{EditEx} is introduced and its effectiveness is evaluated. Users can get the \texttt{EditEx} interface integrated with the SO's existing edit system after installing \texttt{EditEx}. Then, they can suggest edits to posts using \texttt{EditEx} to improve their quality. We measure the effectiveness of \texttt{EditEx} in two ways. 
First, we recruited 20 participants and divided them into two groups. The control group edits posts using the existing SO edit system of SO, whereas the treatment group uses \texttt{EditEx} to edit posts. When both groups complete their edits, we compare the success rates (e.g., fewer rejection ratios). Second, we survey participants after completing their edits and analyze their feedback.


\subsection{Study Design}\label{sec:study-design}
In this section, we first discuss how we recruit our study participants (Section \ref{subsec:participants}). We then explain the formation of control vs. treatment groups (Section \ref{subsec:group-formation}). Next, we discuss the two phases of our study -- task-based evaluation and survey-based feedback collection phase (Section \ref{subsec:execution-plan}).

\subsubsection{Study Participants}
\label{subsec:participants}

We recruit 20 participants who satisfy our constraint (i.e., participants must have editing experience in SO posts) in the following two ways.

\begin{itemize}
    \item Snowball Approach:
    We use convenience sampling to bootstrap the snowball \citep{stratton2021population}. We first contacted a few software developers who are known to us, easily reachable, and working in software companies worldwide. We discuss our study goals and share the survey with them. We then adopt a snowballing method \citep{bi2021accessibility} to disseminate the survey to some of their colleagues with similar experiences. We asked to share the survey with those who could be interested in editing posts and participating in our survey. In this process, we receive information (e.g., email) from 22 participants who show their interest in our study. However, 12 of them finally confirmed their participation.
    
    \smallskip\item Open Circular:
    We post a description of this study and our research goals in the specialized Facebook groups to find potential participants. We target the groups where professional software developers discuss their programming problems. We also use LinkedIn as a research tool to reach potential participants because it is one of the largest professional social networks in the world. We get contacts of 20 participants from this open circular who are willing to participate and satisfy our constraints (e.g., must have editing experience). However, some participants did not respond to us when we contacted them. We finally confirmed \emph{eight} participants in this process.
    
\end{itemize} 
    
\noindent In the end, we were able to recruit a total of 20 developers (12 from snowball + 8 from open circular) who were eligible based on our study constraint. Half of them have 3--5 years, 40\% (8 out of 20) have two years (or less), and 10\% (2 out of 20) have 6--8 years of software development experience. Note that 35\% of participants were from different software industries, and the remaining 65\% of them were from academia. They worked as developers, technical leads, grad students, and faculty members worldwide (e.g., Canada, Germany, Bangladesh).
In our registered report \citep{mondal2020rollbackRegisteredProtocol}, we proposed to recruit 30 participants using a snowball approach. However, we discussed these discrepancies in the number of participants and recruitment approach in Section \ref{sec:deviation}.

\subsubsection{Formation of Control and Treatment Groups}
\label{subsec:group-formation}

As mentioned above (Section \ref{subsec:participants}), we recruit 20 participants for our study. We pick 10 of them for the treatment group and 10 for the control group as follows.

\begin{table}[h]
	\centering
	\captionsetup{justification=centering, labelsep=newline}
	\caption{Experience and profession of control and treatment groups}
	\label{table:control-treatment-groups}
	\resizebox{4.7in}{!}{%
    
    \begin{tabular}{c|c|c|c|c|c|c|c|c|c}
    \toprule
    
    \multicolumn{5}{c|}{\textbf{Control Group}} & \multicolumn{5}{c}{\textbf{Treatment Group}} \\ \midrule
    
    \multicolumn{3}{c|}{\textbf{\begin{tabular}[c]{@{}c@{}} Development \\ Experience (Years)\end{tabular}}}  & \multicolumn{2}{c|}{\textbf{Profession}} & \multicolumn{3}{c|}{\textbf{\begin{tabular}[c]{@{}c@{}}Development \\ Experience (Years)\end{tabular}}} & \multicolumn{2}{c}{\textbf{Profession}} \\ \midrule
    
    \multicolumn{1}{c|}{\textbf{$\leq$ 2}} & \multicolumn{1}{c|}{\textbf{3-5}} & \multicolumn{1}{c|}{\textbf{6-8}} & \multicolumn{1}{c|}{\textbf{Academician}} & \multicolumn{1}{c|}{\textbf{SW Developer}} & \multicolumn{1}{c|}{\textbf{$\leq$ 2}} & \multicolumn{1}{c|}{\textbf{3-5}} & \multicolumn{1}{c|}{\textbf{6-8}} & \multicolumn{1}{c|}{\textbf{Academician}} & \multicolumn{1}{c}{\textbf{SW Developer}} \\ \midrule
    
     4 & 5 & 1 & 6 & 4 & 4 & 5 & 1 & 7 & 3 \\ \bottomrule
    \end{tabular}    

    }
\end{table}

\begin{itemize}
    \item Treatment Group: Each participant in this group was assisted in their editing of SO posts by our developed \texttt{EditEx} tool. However, the participant could also access the standard SO edit system.

    \smallskip\item Control Group: Each participant in this group edited SO posts by using the standard SO edit system only.
\end{itemize}

\noindent Table \ref{table:control-treatment-groups} shows the experience and professions of control and treatment groups. We attempt to equate profession and experience between two groups to minimize the subjective biases. For example, control and treatment groups have participants with low and high software development experience. Also, each group contains participants from academia (e.g., faculty members) and software industries.

\subsubsection{Execution Plan}
\label{subsec:execution-plan}

\begin{table}[t]
	\centering
	\captionsetup{justification=centering, labelsep=newline}
	\caption{Editing tasks to the participants}
	\label{table:edit-tasks}
	\resizebox{4.7in}{!}{%
    \begin{tabular}{c|p{10cm}}
    \toprule
    
    \textbf{Task ID}  & \textbf{Task Description}  \\ \midrule
    
    T1 & Add greeting (e.g., hello, dear, good day) at the beginning of a post with other edits. \\ \midrule
    T2 & Add gratitude (e.g., thank you) at the end of a post with other edits. \\ \midrule
    T3 & Add signature (e.g., user name) at the end of a post with other edits. \\ \midrule
    T4 & Format text (e.g., bold/unbold text, make text italic, format text element as code element or vice versa) with other edits. \\ \midrule
    T5 & Format code (e.g., add/remove space, make lowercase letter to uppercase or vice versa) with other edits. \\ \midrule
    T6 & Remove a code segment or a paragraph of text or change them radically. \\ \midrule
    T7 & Other (e.g., fix grammar \& spelling, text/code modification) \\ \bottomrule

    \end{tabular}
    }
\end{table}
\textbf{\ul{First, we conduct a task-based analysis}} (Section \ref{sec:task-based-analysis}) by asking each participant (control + treatment group) to edit ten posts. However, we set a task list and asked each participant to suggest edits to posts following the task list. Table \ref{table:edit-tasks} lists the tasks. Tasks T1--T6 were set based on the identified reasons that might cause rejections. For example, the treatment and control groups were asked to add gratitude (e.g., thank you) associated with other edits. In this case, \texttt{EditEx} warns the treatment group against adding such gratitude. Therefore, participants of the treatment group proceed with the remaining edits and avoid adding gratitude.
On the other hand, the control group did not receive such a warning. Thus, they suggested edits with gratitude. We attempt to see the effectiveness of \texttt{EditEx} in preventing the commonly rejected reasons while suggesting edits from tasks T1--T6. However, the control group might get more rejections for tasks T1--T6. On the contrary, such tasks favor the treatment group. Such a scenario can exaggerate \texttt{EditEx}'s effectiveness in reducing edit rejections. Therefore, we also ask the participants to edit posts arbitrarily (e.g., Table \ref{table:edit-tasks}, T7). That means participants suggest edits to posts that are not related to any rejection reasons to limit the bias of this study.

We circulated the editing guidelines of SO to each participant of the control and treatment groups. We asked participants to follow the guidelines in suggesting edits. After suggesting edits, participants wait until they get the decision (rejected/accepted) on those suggested edits from the edit reviewers. However, users with a reputation score $\geq 2K$ can edit posts instantly. Those edits are neither added to the review queue nor reviewed by experts. However, a rollback can reject their edits. Unfortunately, such a rollback even may take a few months. We thus ask participants with a reputation score $\geq 2K$ to create a new account. Edits suggested by their new account undergo an expert review. This decision also confirms the same privilege level of all the participants. However, our target was to quickly get decisions (rejected/accepted) on the suggested edits and avoid undecided edits.

\noindent\textbf{\ul{Second, we conduct an online survey}}  (Section \ref{sec:survey}) to listen to the participants about their experience in suggesting edits to SO posts with/without our tool \texttt{EditEx}.
Kitchenham and Pfleeger~\citep{kitchenham2008personal} suggest considering six main steps for a personal opinion survey. They are -- setting survey objectives, designing the survey, developing the survey instrument (i.e., the questionnaire), evaluating the survey instrument, and obtaining and analyzing data. We primarily follow their guidelines to survey participants. However, we also consider the guidance and ethical issues from the established best practices~\citep{groves2011survey, singer2002ethical}. For example, we take participants' consent before starting the survey. Besides, we confirm to participants that their provided information must be treated confidentially. Our survey includes different types of questions (e.g., multiple-choice, free-text answers). However, we inform about the estimated time (i.e., approximately 10 minutes) required to complete the survey to the participants. Our survey comprises the parts as follows.

\begin{itemize}
    \smallskip\item \textbf{Consent and Prerequisite.} This part confirms participants' consent to participate in this survey and agreement to process their data.

    \smallskip\item \textbf{Participants' Information.} In this part, we collect participants' information such as experience, current profession, organization, country, and editing experience in SO posts.
    

    \smallskip\item \textbf{Workload Assessment.} This section assesses the cognitive workload of the control and treatment groups in suggesting edits to SO posts. We leverage the NASA Task Load Index (TLX) (non-weighted) to estimate subjective workload \citep{cao2009nasa, hart1988development, noyes2007self, sharek2011useable}. In particular, we assess how much effort participants had to exert mentally and physically to use \texttt{EditEx} and the standard edit system of SO. Participants were asked to rate their scores on an interval scale ranging from low ($1$) to high ($10$) \citep{memarian2011work} in the following six dimensions -- (1) mental demand, (2) physical demand, (3) temporal demand, (4) effort, (5) performance, and (6) frustration.

        
    
        
        
        
        

    \smallskip\item \textbf{Usefulness Analysis.} In this section, we measure the participants' confidence in suggesting edits using \texttt{EditEx} (treatment group) and SO edit system (control group). In addition, we ask the treatment group to rate the usefulness of the suggestions of \texttt{EditEx}. In particular, we ask the following two questions to the treatment group participants and employ a 5-point Likert scale (i.e., 1--5) to estimate their consent \citep{joshi2015likert, vagias2006likert}.
    
        \smallskip\begin{enumerate}
            \item How useful did you find the suggestions from \texttt{EditEx}? (\emph{5-point Likert scale})
            \item How confident were you to follow the \texttt{EditEx} suggestions? (\emph{5-point Likert})
        \end{enumerate} 
        
    \smallskip\noindent We ask the following question to measure the confidence level of the participants in the control group.
        
        \begin{enumerate}[start=3]
            \item How confident are you to edit posts using the SO editing system? (\emph{5-point Likert scale})
        \end{enumerate}

    \smallskip\item\textbf{Suggestions to Improve \texttt{EditEx}.} Finally, we seek participants' recommendations to improve the effectiveness of \texttt{EditEx}. We ask them the question as follows.
    
        \smallskip\begin{enumerate}
            \item What are your recommendations to further improve \texttt{EditEx}? (\emph{Text})
        \end{enumerate}

\end{itemize}

\noindent We added the \emph{survey form} and its \emph{responses} in anonymized CSV form in our replication package~\citep{replicationPackage}.

\subsection{Results from the User Study on Editing of SO Posts}
\label{sec:task-based-analysis}

Participants were asked to complete ten edits. However, several participants (especially from the control group) could not edit ten posts due to three main challenges.

\begin{enumerate}

    \item SO's edit queue often remains full, and thus participants could not edit posts according to their schedule.
    
    \item Participants could not edit many posts (e.g., more than three) simultaneously. SO restricts them from suggesting further edits before receiving a decision (rejected/accepted) on the pending ones.
    
    \item SO does not allow its users to edit for a period when consecutive edits are being rejected.
    
\end{enumerate}

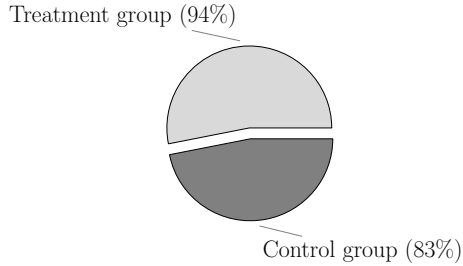
\begin{figure}[!htb]
	\centering
	\resizebox{2.4in}{!}{
    \begin{tikzpicture}
    \pie[explode=0.2, text=pin, sum = auto, color={black!15, black!50}, number in legend] 
        { 94/\Huge{Treatment group (94\%)},
          83/\Huge{Control group (83\%)}
          }
    \end{tikzpicture}
    }
	\caption{Task completion of treatment and control groups.}
	\label{fig:test-completion-ratio}
\end{figure}

\noindent Fig. \ref{fig:test-completion-ratio} shows the task completion ratio of control and treatment groups. As mentioned above, we asked each group to edit 100 posts (10 for each). However, participants from the control group were able to edit 83 posts in total (i.e., completion ratio 83\%), whereas the treatment group edited 94 posts.
\begin{figure}[!htb]
	\centering
	\resizebox{3.3in}{!}{
    \begin{tikzpicture}
    \pie[explode=0.2, text=pin, sum = auto, color={black!15, black!50}, number in legend] 
        { 16/\Huge{Treatment group (16\%)},
          65.1/\Huge{Control group (65.1\%)}
          }
    \end{tikzpicture}
    }
	\caption{Rejection ratio of treatment and control groups.}
	\label{fig:rejection-ratio}
\end{figure}
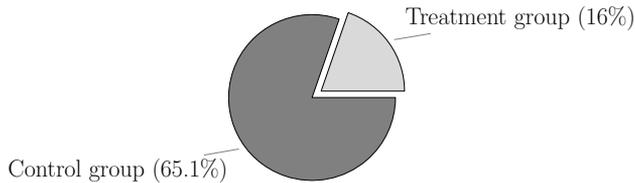
We then attempt to see the rejection ratio of suggested edits. First, we collect information from each participant (treatment \& control group) when all of their suggested edits got decisions (e.g., rejected/accepted). In particular, we asked how many edits they suggested and how many of them got rejected. We also collect their editing details to examine whether they follow the given task list (e.g., Table \ref{table:edit-tasks}) or not. Then we count the total number of suggested and rejected edits of the treatment and control groups. Finally, we calculate the rejection ratio of each group.

As shown in Fig. \ref{fig:rejection-ratio}, the rejection ratio of the edits suggested by the treatment group is only 16\% (15 out of 94). On the contrary, such a statistic is 65.1\% (54 out of 83) for the control group. Overall, the percentage of rejected edits who used \texttt{EditEx} is about 49\% lower than those who used the standard editing system of SO. Such a finding gives us a preliminary validation that our tool helps users to prevent their suggested edits from being rejected.

\begin{figure}[!htb]
	\centering
	\resizebox{3in}{!}{
    \begin{tikzpicture}
    \pie[explode=0.2, text=pin, sum = auto, color={black!15, black!50}, number in legend] 
        { 14.1/\Huge{Treatment group (14.1\%)},
          26.1/\Huge{Control group (26.1\%)}
          }
    \end{tikzpicture}
    }
	\caption{Rejection ratio of \textbf{T7} between treatment and control groups.}
	\label{fig:rejection-ratio-t7}
\end{figure}
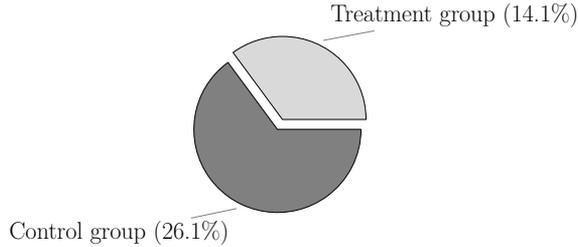

Tasks T1--T6 were set based on our identified rejection reasons. However, \texttt{EditEx} alerts the treatment group participants while suggesting T1--T6. Such alerts might help them to avoid edits that cause rejections. On the contrary, the control group did not receive any alerts from the SO edit system while suggesting T1--T6. Therefore, the overall edit rejection ratio of the control group is much higher than the treatment group. We thus attempt to compare the rejection ratio of task T7 (i.e., free-form editing) between control and treatment groups. According to our analysis, the control group suggests 23 free-form edits. Among them, six were rejected. Since \texttt{EditEx} prevents T1-T6, the treatment group could not suggest T1-T6. Therefore, we consider all of their tasks as T7. However, we found that they suggested two edits out of task T7 (i.e., T1--T6). Among the remaining 92, 13 edits were rejected. As shown in Fig. \ref{fig:rejection-ratio-t7}, the rejection ratio of the control group for T7 is 26.1\%. On the other hand, such ratio for the treatment group is 14.1\%. 
Therefore, \texttt{EditEx} not only assists users in avoiding edits that are usually rejected but also assists them in conducting regular (i.e., free-form) edits.

\subsection{Results from Survey of User Study Participants}
\label{sec:survey}

We received 20 valid survey responses (10 treatment + 10 control). We report the survey responses as follows.

\begin{itemize}
    \item \textbf{Workload assessment}  during the completion of the editing tasks, 
    \item \textbf{Usefulness ratings} of \texttt{EditEx} suggestions,  and 
    \item \textbf{Improvement suggestions} by the study participants for \texttt{EditEx}.
\end{itemize}

\subsubsection{Workload Assessment During Edit Task Completion}
\label{subsec:workload}

\begin{figure}[t]
\centering
        \resizebox{2.1in}{!}{%

        \begin{tikzpicture}
        \begin{axis}[xmin=0,width=6cm,height=6cm,
          ylabel = {Average NASA Task Load Value},
          xtick = {1, 2},
          xticklabels = {Treatment, Control},
          ]
        \addplot+[boxplot, 
                  /pgfplots/boxplot/hide outliers, 
                  boxplot/draw direction = y,
                  mark = *,
                  boxplot,
                  fill,fill opacity=0.0,
                  black]
        table[row sep=\\,y index=0] {
        data\\
        2.5\\
        2\\
        1.833333333\\
        1.333333333\\
        3.166666667\\
        2.666666667\\
        4.333333333\\
        4.166666667\\
        5\\
        3.666666667\\
        };
        
        \addplot+[boxplot, 
                  /pgfplots/boxplot/hide outliers, 
                  boxplot/draw direction = y,
                  mark = *,
                  boxplot,
                  fill,fill opacity=0.2,
                  black]
        table[row sep=\\,y index=0] {
        data\\
        7.83333333\\
        7.333333333\\
        6\\
        6\\
        7.833333333\\
        6\\
        4.333333333\\
        6.333333333\\
        7\\
        6\\
        };
        
        \end{axis}
        \end{tikzpicture}
        }
\caption{Cognitive workload in editing SO posts using \texttt{EditEx} vs. SO's standard edit system using NASA TLX.}
\label{fig:nasa-tlx-workload}
\end{figure}
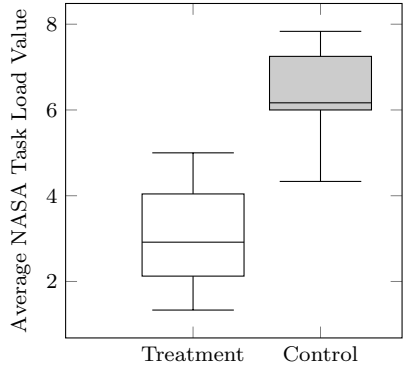

Fig. \ref{fig:nasa-tlx-workload} shows the box plots of the NASA TLX cognitive workload score on a scale of ten. We compute the average workload of each participant. We first sum up the ratings of each of the six dimensions (e.g., mental demand) and then divide it by the number of dimensions. In particular, we use the equation to compute the average workload of each participant as follows.

\begin{equation}
A_{wl} = \frac{1}{D_T}\Bigg[\sum_{i=1}^{D_{T}} R_{i}\Bigg]
\end{equation} where $R_i$ denotes the rating (1--10) of i\textsuperscript{th} dimension, $D_T$ represents the total dimensions (here, $D_T = 6$). 

Fig. \ref{fig:nasa-tlx-workload} shows the box plots that represent the average cognitive workload of each participant from the treatment and control groups. We see that the median subjective workload for the treatment group is about half that of the control group. That is, \texttt{EditEx} strongly supports the SO standard editing system to reduce the users' workload required to suggest edits to posts. 
We then attempt to find whether the workload difference between treatment and control groups is statistically significant. We use the \emph{Mann-Whitney-Wilcoxon} statistical significance test \citep{mcknight2010mann} and find a statistical significance \emph{p-value} (p-value $\simeq$ $0.0$ $<$ $0.05$). We also use \emph{Cliff's delta} test \citep{macbeth2011cliff} to determine the effect size and find a large \emph{effect size} (Cliff's d $= -0.97$ (large)) with $95\%$ confidence. Given this evidence, \texttt{EditEx} helps the users to edit posts by significantly reducing their workload.

\subsubsection{Usefulness of \texttt{EditEx} Suggestions}
\label{subsec:usefulnessEditEx}

Table \ref{table:effectiveness-analysis} shows the participants' assessment of the effectiveness of \texttt{EditEx} and the SO edit system. We see that participants find the suggestions of \texttt{EditEx} \emph{influential} (3.41 $\leqslant$ score $\leqslant$ 4.20) in avoiding the potential rejections. The Likert score (i.e., 4.0) also shows that they were \emph{confident} to follow the suggestions given by \texttt{EditEx}. When we asked the reason behind their confidence level, one participant responded that \emph{\texttt{EditEx} suggests the common reasons behind the unsuccessful attempts of edit. These suggestions help to identify those and fix them}. On the contrary, the participants who use the SO edit system (i.e., control group) were \emph{moderately confident} (2.61 $\leqslant$ score $\leqslant$ 3.40) in suggesting their edits. To explain the reasons, one participant stated that \emph{I was not sure whether my edits are good or bad, so my confidence was low. I cannot understand much from the reviewers' comments why my edits were actually rejected. The reasons were too general}. Such findings indicate that \texttt{EditEx} is not only able to provide valuable suggestions but also make users more confident in suggesting edits.

\begin{table}[t]
	\centering
	\captionsetup{justification=centering, labelsep=newline}
	\caption{Effectiveness analysis of \texttt{EditEx} and SO edit system}
	\label{table:effectiveness-analysis}
	\resizebox{4.7in}{!}{%
    \begin{tabular}{p{6cm}|c|c|c}
    \toprule
    
    \multicolumn{1}{c|}{\textbf{Questions}}  & \textbf{Target Group} & \textbf{Mean Value} & \textbf{Interpretation} \\ \midrule
    
    How useful did you find the suggestions from \texttt{EditEx}? & Treatment & 4.1 & Influential \\ \midrule
    How confident were you to follow the \texttt{EditEx} suggestions? & Treatment & 4.0 & Confident\\ \midrule
    How confident are you to edit posts using the SO editing system? & Control & 3.1 & Moderate\\ \bottomrule
    \end{tabular}
    }
\end{table}


\subsubsection{Recommendations for EditEx Improvements}\label{sec:editex-improvement}


    

We analyzed the recommendations of all the participants and summarized them into three categories. We see that participants recommended --  (1) enhancing existing functionalities, (2) improving Graphical User Interface (GUI), and (3) notification \& installation system. We discuss their recommendations below.

\vspace{2mm}
\noindent \textbf{Enhance Functionality} \vspace{-2mm}

\begin{itemize}

    \item Besides suggesting the rejection reasons, \texttt{EditEx} could also estimate a score based on the quality of the edit. 
    
    \item \texttt{EditEx} can be enhanced by adding a few features related to natural language processing, such as identifying incorrect spelling, sentence complexity.
    
    \item \texttt{EditEx} should detect minor changes (e.g., adding an article) that do not significantly improve the quality of the posts. 
    
    \item Participants also suggested enhancing the capability of \texttt{EditEx} in such a way that it can identify more potential reasons that might cause rejection.
    
    \item Participants recommend paraphrasing the notification sentences (e.g., edits may get rejected due to low reputation) to convey them more positively.

\end{itemize}

\noindent \textbf{Improve Graphical User Interface} \vspace{-2mm}

\begin{itemize}
    
    \item A few participants recommend improving the GUI of \texttt{EditEx}. For example, the notification system of the potential rejection reasons could be more appealing. In addition, \emph{Suggest Me} button should appear beside the edit window to avoid scrolling.

\end{itemize}

\noindent \textbf{Add Notification \& Improve Installation System} \vspace{-2mm}

\begin{itemize}
    
    \item One of the main barriers to suggesting edit is that the queue remains full most of the time. Participants thus suggest that EditEx should notify them when the queue becomes free to suggest edits to avoid frequent manual checking.
    
    \item \texttt{EditEx} uses Tampermonkey to add userscripts for integrating it into the SO edit system. They appreciate it since Tampermonkey is popular and easy to use. However, they suggest deploying \texttt{EditEx} as a standalone browser plug-in in future. 

\end{itemize}

\smallskip\begin{tcolorbox}[colframe=black!50,colback=white,left=0pt,right=1pt,top=1pt,bottom=1pt, boxrule=1pt,arc=1pt, width=4.7in, center]

    \textbf{Summary of $RQ_2$ and hypothesis testing:} Results from the user study on the editing of SO posts show that \texttt{EditEx} can support the SO edit system to prevent overall 49\% of rejected edits, including the commonly rejected ones. However, it can prevent 12\% rejections even in regular free-form edits. On the other hand, results from the survey of user study participants show that users who use \texttt{EditEx} found its suggestions influential while editing posts. Therefore, they were confident to follow the suggestions. As a result, \texttt{EditEx} can reduce the cognitive workload in half compared to the SO edit system. 
    Such findings \emph{reject} the null hypothesis ($H_2$), which hypothesized that the \texttt{EditEx} tool is ineffective in helping users avoid rejected edits.  
    
\end{tcolorbox}

\section{Discussions}
\label{sec:discussion}

In this section, we first explain the importance of the features used in our machine learning models (Section \ref{subsec:featureRank}). We then discuss the reasons behind the misclassifications of the machine learning models (Section \ref{subsec:misclassification-analysis}). Finally, we discuss the implications of our study findings and the developed \texttt{EditEx} tool in Section \ref{subsec:implications}.

\subsection{Ranking of Features in the Machine Learning Models}
\label{subsec:featureRank}

\begin{figure}[!htb]
\centering
   	\pgfplotstableread{
                1	0.039713704	0.610002421
                2	0.031210201	0.606805541
                3	0.019863832	0.1361536
                4	0.016820502	0.023263341
                5	0.013169681	0.008802344
                6	0.007769412	0.006311943
                7	0.002785819	0.004563579
                8	0.001992525	0.003255283
                9	0.001614938	0.001110599
                10	0.001553978	0.000490381
                11	0.001435972	0.000162124
                12	0.001377221	0.000108413
                13	0.000218634	9.21E-06
                14	0.000218313	7.84E-08
                15	0	0

    	}\datatable
       \subfloat[Information gain]{
      \label{fig:info-gain}
      \resizebox{2.3in}{!}{%
      \begin{tikzpicture}
        	\begin{axis}[
        	ylabel=Score (Information gain),
        	xtick=data,
        	xticklabels={text change,
                        reputation,
                        code change,
                        gratitude,
                        deface post,
                        text format,
                        signature,
                        duplication,
                        greetings,
                        reference modification,
                        complete change,
                        inactive link,
                        status,
                        code format,
                        deprecation
                        },
        	x tick label style={rotate=90,anchor=east},
        	enlarge y limits=false,
        	enlarge x limits=0.07,
        	ymin=0,ymax=1,
        	ybar,
        	bar width=0.2cm,
        	width=3in,
        	height = 2in,
        	ytick={0,0.1,0.2,0.3,0.4,0.5,0.6,0.7,0.8,0.9,1},
            yticklabels={0,0.1,0.2,0.3,0.4,0.5,0.6,0.7,0.8,0.9,1},
        	ymajorgrids=false,
        	major x tick style = {opacity=0},
        	minor x tick num = 1,    
        	minor tick length=1ex,
        	legend style={
         	legend pos=north east,
        	legend cell align=left
            },
        	]
        	\addplot[draw=black!80, fill=black!80] table[x index=0,y index=2] \datatable;

        	\end{axis}
    	\end{tikzpicture}
     	}
     	}
      \subfloat[SHAP feature importance]{
      \label{fig:shap-score}
      \resizebox{2.3in}{!}{%
      \begin{tikzpicture}
        	\begin{axis}[
        	ylabel=Mean($|$SHAP value$|$),
        	xtick=data,
        	xticklabels={reputation,
                        code change,
                        gratitude,
                        text change,
                        text format,
                        deface post,
                        signature,
                        status,
                        duplication,
                        reference modification,
                        greetings,
                        inactive link,
                        code format,
                        complete change,
                        deprecation
                        },
        	x tick label style={rotate=90,anchor=east},
        	enlarge y limits=false,
        	enlarge x limits=0.07,
        	ymin=0,ymax=0.10,
        	ybar,
        	bar width=0.2cm,
        	width=3in,
        	height = 2in,
        	ytick={0,0.01,0.02,0.03,0.04,0.05,0.06,0.07,0.08,0.09,0.1},
            yticklabels={0,0.01,0.02,0.03,0.04,0.05,0.06,0.07,0.08,0.09,0.1},
        	ymajorgrids=false,
        	major x tick style = {opacity=0},
        	minor x tick num = 1,    
        	minor tick length=1ex,
        	legend style={
         	legend pos=north east,
        	legend cell align=left
            },
        	]
        	\addplot[draw=black!80, fill=black!80] table[x index=0,y index=1] \datatable;

        	\end{axis}
    	\end{tikzpicture}
     	}
     	}
     	
\caption{Ranking of features.}
\label{fig:reature-ranking}
\end{figure}
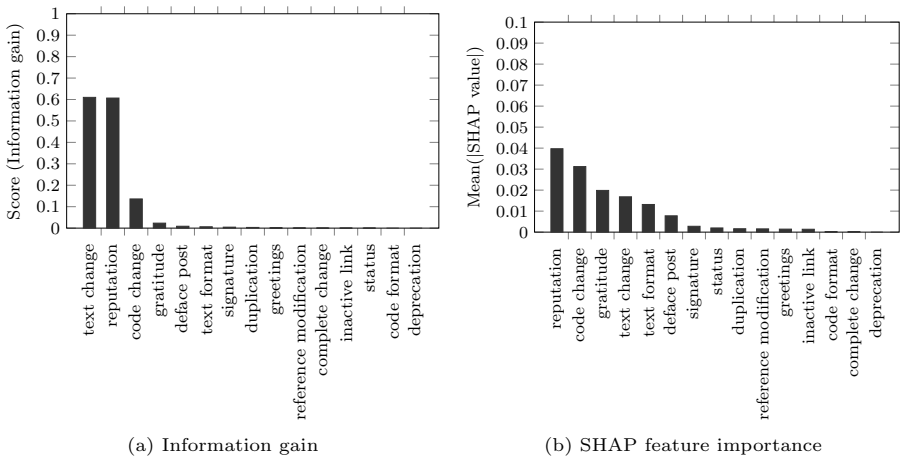

In Section \ref{subsec:featureExtraction}, we discussed several features used to develop our machine learning models in predicting whether an edit will be rejected or accepted. However, we do not know which features are more important in machine learning classifiers than others in differentiating rejected and accepted edits. To find the important features, we thus attempt to rank our features using two popular measures as follows.

\smallskip\noindent\textbf{Information Gain.} 
We attempt to determine which features are more robust than others for discriminating between rejected and accepted edits. We thus employ an information gain-based feature ranking technique because it can estimate the discrimination power of each of the given features. In information theory, the information gain of a random variable is the change in information entropy between an initial state and a state that takes some information \citep{TUT-Saha-2013}. Therefore, the information gained from a particular attribute in classifying rejected edits is as follows.

\begin{equation}
	Info\hspace{0.5mm}Gain(C, a{_i}) = H(C) - H(C|a{_i})
\end{equation} where $C$ represents a particular class (i.e., rejected/accepted), $a{_i}$ denotes the attribute, and $H$ denotes information entropy. 

Fig. \ref{fig:info-gain} shows the information gain of our selected features. We see that changes in text and reputation score have the highest information gain. It means that they might discriminate the edits more accurately than others. Changes in code and addition/removal of gratitude have the next highest information gain. However, the information gains of the remaining features are minimal. Therefore, those features contribute less to the machine learning models to classify the rejected edits from the accepted ones.

\smallskip\noindent\textbf{SHapley Additive exPlanations (SHAP) Feature Importance.} 
The SHAP value is the average marginal contribution of a feature towards the model’s prediction across all possible combinations of features \citep{molnar2020interpretable}. It shows whether a feature value can increase a model’s prediction over a random baseline \citep{lundberg2020local}. However, the idea behind SHAP feature importance is that features with larger absolute SHAP values are more important than others. SHAP values can be calculated for any tree-based model. We calculate the  SHAP values to rank features from the Random Forest model.

Fig. \ref{fig:shap-score} shows the SHAP values of our selected features. We see that the top four features are the same as when using the information gain-based feature ranking technique. According to SHAP values, code change has the second-highest capability of identifying rejected edits following the reputation score. However, the addition/removal of gratitude has more power to discriminate rejected edits from accepted ones over changes in texts.
We then attempt to see the actual effect of our selected features on the model's performance. We thus remove the low-ranked features (according to SHAP values) one by one and evaluate the performance of our model. Table \ref{table:effect-individual-feature} shows the experimental results. According to the experiment, the model's performance gradually degrades when we remove features one by one. For example, overall accuracy decreases 2\% when we remove deprecation.
Interestingly, we see a slightly higher performance when we keep the top six features than the top seven. Then the performance again decreases after removing features. However, the overall accuracy drops from 69.8\% to 61.3\% when we remove all other features except the top four.

\begin{table}[!htb]
	\centering
	\captionsetup{justification=centering, labelsep=newline}
	\caption{Effect of individual feature on predicting rejected edits \\
	\footnotesize{(f\textsubscript{1}: reputation, f\textsubscript{2}: code change, f\textsubscript{3}: gratitude, f\textsubscript{4}: text change, f\textsubscript{5}: text format, f\textsubscript{6}: deface post, f\textsubscript{7}: signature, f\textsubscript{8}: status, f\textsubscript{9}: duplication, f\textsubscript{10}: reference modification, f\textsubscript{11}: greetings, f\textsubscript{12}: inactive link, f\textsubscript{13}: code format, f\textsubscript{14}: complete change, f\textsubscript{15}: deprecation)}}
	
	\label{table:effect-individual-feature}
	\resizebox{4in}{!}{%

    \begin{tabular}{l|c|c|c|c|c|c|c}
    \toprule
    \multirow{2}{*}{\textbf{Features}} & \multicolumn{3}{c|}{\textbf{Rejected}} & \multicolumn{3}{c|}{\textbf{Accepted}} & \multirow{2}{*}{\textbf{Accuracy}} \\

        & \textbf{Precision} & \textbf{Recall} & \textbf{F1-Score} & \textbf{Precision} & \textbf{Recall} & \textbf{F1-Score} &     \\ \midrule
        
    f\textsubscript{1} -- f\textsubscript{15} & 69.1\% & 71.2\% & 70.1\% & 70.5\% & 68.4\% & 69.5\% & 69.8\% \\ \midrule
    
    f\textsubscript{1} -- f\textsubscript{14} & 68.3\% & 65.9\% & 67.1\% & 67.4\% & 69.7\% & 68.5\% & 67.8\% \\ \midrule
    
    f\textsubscript{1} -- f\textsubscript{13} & 65.9\% & 71.6\% & 68.6\% & 69.2\% & 63.2\% & 66.1\% & 67.4\% \\ \midrule
    
    f\textsubscript{1} -- f\textsubscript{12} & 65.5\% & 72.9\% & 69.0\% & 69.8\% & 61.9\% & 65.6\% & 67.4\% \\ \midrule
    
    f\textsubscript{1} -- f\textsubscript{11} & 66.7\% & 69.0\% & 67.8\% & 68.2\% & 65.8\% & 67.0\% & 67.4\% \\ \midrule
    
    f\textsubscript{1} -- f\textsubscript{10} & 62.9\% & 74.7\% & 68.3\% & 69.2\% & 56.3\% & 62.1\% & 65.4\% \\ \midrule
    
    f\textsubscript{1} -- f\textsubscript{9} & 63.6\% & 73.4\% & 68.2\% & 68.9\% & 58.4\% & 63.2\% & 65.9\% \\ \midrule
    
    f\textsubscript{1} -- f\textsubscript{8} & 61.5\% & 73.8\% & 67.1\% & 67.6\% & 54.1\% & 60.1\% & 63.9\% \\ \midrule
    
    f\textsubscript{1} -- f\textsubscript{7} & 62.1\% & 71.6\% & 66.5\% & 66.8\% & 56.7\% & 61.4\% & 64.1\% \\ \midrule
    
    f\textsubscript{1} -- f\textsubscript{6} & 64.4\% & 71.2\% & 67.6\% & 68.1\% & 61.0\% & 64.4\% & 66.1\% \\ \midrule
    
    f\textsubscript{1} -- f\textsubscript{5} & 61.3\% & 72.1\% & 66.3\% & 66.5\% & 55.0\% & 60.2\% & 63.5\% \\ \midrule
    
    f\textsubscript{1} -- f\textsubscript{4} & 59.4\% & 70.3\% & 64.4\% & 64.0\% & 52.4\% & 57.6\% & 61.3\% \\ \bottomrule

    \end{tabular}
    }
\end{table}

\begin{figure}[htb]
	\centering
	\includegraphics[width=4in]{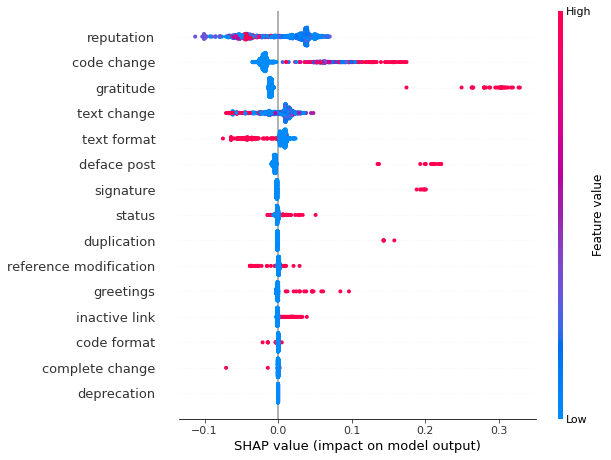}
	\caption{Feature importance using bee swarm plot (Random Forest model).}
	\label{fig:shap-exaplanation}
\end{figure}

\subsection{Analysis of Trained Machine Learning Model \& Its Misclassifications}
\label{subsec:misclassification-analysis}

Machine learning models can produce accurate/inaccurate predictions. However, their \emph{black box} nature might prevent their easy adoption and enhancement by others. The SHAP \citep{lundberg2017unified} is a popular model interpretation framework to interpret the classification/misclassification results of the model. In our experiment, we conduct a binary classification where the \emph{rejected edits} was considered as the \emph{positive} class and \emph{accepted edits} as the \emph{negative} class. Thus, our models attempt to predict the rejected edits by default. Therefore, a positive SHAP value indicates an increase in our models’ prediction of positive class and vice versa. Fig. \ref{fig:shap-exaplanation} shows the importance of our selected features using a bee swarm plot from our random forest model. The bee swarm plot visualizes the SHAP value of a feature from each of the training instances on the x-axis. On the y-axis, it sorts all features in descending order according to their sum of SHAP values. The blue color indicates a low feature value, whereas the red indicates a high feature value in our plot. We see that reputation is the most important feature according to our random forest model. That is, community trust, which is estimated by reputation score, is an important predictor of the acceptability of suggested edits. We note that this feature with true response often leads to negative SHAP values, which indicates an increased prediction towards edit acceptance. That is, edits suggested by users with a high reputation score have a higher chance of being accepted and vice versa. 
We also analyze our dataset for further insights on this. In particular, we conduct a comparative analysis of the reputation scores between users with rejected and accepted edits. We find the median reputation score of users whose suggested edits were rejected is $3428$. On the contrary, such a score of users whose suggested edits were accepted is more than double (i.e., $7660$). We then attempt to see whether the difference is statistically significant or not. We use the \emph{Mann-Whitney-Wilcoxon} statistical significance test and find a statistical significance \emph{p-value} (p-value $\simeq$ $0.0$ $<$ $0.05$). We also use \emph{Cliff's delta} test to determine the effect size and find a medium \emph{effect size} (Cliff's d $= -0.34$ (medium)) with $95\%$ confidence. Given this evidence, low-reputed users are less trusted by the community, and thus their edits can be rejected.

The second most important feature is the code change. Code change with true response often leads to positive SHAP values, which indicates an increased prediction towards edit rejections. It refers that suggested edits get rejected when users change the code much. On the contrary, text change with true response leads to negative SHAP values, which means that significant changes in texts are acceptable. Similarly, we see that the true response of text \& code format, reference modification, and complete change of posts lead to negative SHAP values, which indicates an increased prediction towards edit acceptance. On the other hand, the true response of gratitude, deface post, signature, status, greetings, and inactive links lead to positive SHAP values, which indicates an increased prediction towards edit rejections. However, according to the SHAP visualization, a few features such as reputation, code \& text change, status, and reference modifications confuse our model. For example, reference modification's true response leads to positive and negative SHAP values. Therefore, those features might cause misclassifications of our model.

\subsection{Implications of Study Findings}
\label{subsec:implications}

The findings from our study and the tool \texttt{EditEx} can guide the following major stakeholders in crowd-sourced knowledge-sharing platforms that use collaborative editing features: \begin{inparaenum}
\item \textbf{forum designers} to improve the edit system, 
\item \textbf{forum users} to guide their edit behavior, and 
\item \textbf{researchers} to study and improve collaborative editing support in crowd-shared platforms.
\end{inparaenum} We discuss the implications below.

\smallskip\noindent\textbf{Forum Designers.} The quality assurance of shared content is paramount for the usefulness and popularity of a crowd-shared knowledge-sharing platform like SO. While the editing of content allows users to suggest improvements in quality, the lack of proper guidance to the user can lead to unnecessary rejections of the suggested edits. SO can use our developed machine learning model and the \texttt{EditEx} tool to offer on-demand and context-aware edit fix recommendations to the users. As we observed in Section \ref{sec:survey}, the SO users that used \texttt{EditEx} had significantly less number of rejections compared to the users that did not use \texttt{EditEx} tool. The SO edit assessment queue is set up to ensure novice SO users (with less than 2K reputation) can not make bad edits. Many users in SO fall under this novice category, but their inputs to edits are equally as important as the non-novice users. However, as we noted, the SO edit queue can be often congested with many suggested edits that it could take a disproportionate time for an edit to get reviewed. Even after the review, many suggestions can be rejected due to trivial edits like undesired text formatting. A tool like \texttt{EditEx} can help SO users by reducing such trivial edits, which then can ultimately help SO and its expert edit reviewers with less rejection, which in turn can reduce the workload on the edit queue. 

\smallskip\noindent\textbf{Forum Users.} Interactive browser plug-ins like \texttt{EditEx} can warn forum users of potential edits that could be rejected. Thus \texttt{EditEx} can improve the confidence of SO users during the editing. Indeed, as we reported in Section \ref{sec:survey}, the study participants were more confident while using \texttt{EditEx} than using SO editing system (average confidence of 4 while using \texttt{EditEx} vs 3.1 while using SO). The SO users had only 16\% of their edits rejected while using \texttt{EditEx}, while the rejection rate was 65\% while using SO. While a rejection ratio of 65\% for SO could be biased due to our choice of editing task, the much lower rejection rate while using \texttt{EditEx} does indeed highlight that SO users can benefit from a simple tool like \texttt{EditEx}. \texttt{EditEx} is very easy to install. Given that \texttt{EditEx} is simply a browser plug-in that SO users can easily install, we hope that \texttt{EditEx} will be accepted by the wider SO community.

Indeed, our developed classifiers can be used to detect rejected edit reasons in SO automatically. It can extend current tools and techniques that predominantly use contents from suggested edits to recommend editing suggestions (e.g., see the works of Chen et al.~\citep{Chen-DeepLearningCollaborativeEdit-CSCW2017,Chen-DeepLearningCollaborativeEditing-MSR2013}). The tool \texttt{EditEx}, with further modification (as suggested by our study participants in Section \ref{sec:editex-improvement}), can be influential in the reduction of rejected edit reasons in SO and can improve the overall satisfaction of SO users. In the long term, the tool can promote better content because users will be more motivated. Such high quality contents then can offer better content and recommendation support for tools and techniques that focus on the quality of contents shared in SO~\citep{Zhang-AreCodeExamplesInForumReliable-ICSE2018,Ya-WantAGoodAnswer-Arxiv2013,Hudson-TriggerForClarificationRequestsForums-VLHCC2015,Masud-InsightUnresolvedQuestionsSO-MSR2015,Agichtein-FindingHighQualityContents-WSDM2008,Mondal-SOIssueReproducability-MSR2019}, the suite of tools and techniques developed to detect and recommend quality posts~\citep{Ponzanelli-ClassifyQualityForumQuestion-QSIC2014,Ponzanelli-ImproveLowQualityPostDetect-ICSME2014,Ya-DetectHighQualityPosts-JIS2015,Harper-PredictorAnswerQuality-CHI2008,Li-AnswerQualityPredictions-WWW2015,Calefato-HowToAskForTechnicalHelp-IST2018}.
 

\smallskip\noindent\textbf{Researchers.} The quality of knowledge shared in SO is important because developers worldwide now rely on this shared knowledge. Indeed, knowledge shared in SO can support diverse activities like bug fixing, feature enhancement, API selection, and documentation~\citep{Uddin-OpinerReviewAlgo-ASE2017,Uddin-OpinerReviewToolDemo-ASE2017,Uddin-APIAspectMining-TechReport2017,Uddin-OpinionSurvey-TSE2019,Uddin-OpinionValue-TSE2019,Uddin-OpinerAPIUsageScenario-IST2020,Uddin-OpinerAPIUsageScenario-TOSEM2020,Chakraborty-NewLangSupportSO-IST2021}. This sharing of knowledge is important, because official software documentation can often be lacking~\citep{Uddin-HowAPIDocumentationFails-IEEESW2015,Khan-DocSmell-SANER2021}. However, the editing of content is a voluntary activity. SO users can be demotivated to produce quality edits if they become frustrated due to unnecessary/unwanted rejection of their suggested edits. Tools like \texttt{EditEx} or proactive policy assurance by Chen et al.~\citep{Chen-DataDrivenPolicyPostQuality-CSCW2018,Chen-DeepLearningCollaborativeEdit-CSCW2017} can help SO users with suggestions to improve their edits. The positive survey responses of our tool \texttt{EditEx} show the potential of deploying the tool in SO. Future research can contribute by including more features into \texttt{EditEx} and by conducting more studies to learn how SO users can further benefit from such tools. Such findings can promote quality contents, which then can support all development tasks that rely on SO, as noted above.

\section{Threats to Validity}
\label{sec:threat}

Threats to internal validity relate to experimental errors and biases \citep{tian2014automated}. We asked participants to suggest edits using \texttt{EditEx} (treatment group) and SO's standard edit system (control group). The suggested edits were either accepted or rejected by the expert review. However, the accepted edits could be rejected later by a rollback that might affect the edit rejection ratio. We thus further analyzed how many accepted edits were rejected by rollbacks in our dataset. However, such a statistic was less than 1\% in our dataset. Thus, it might not affect our results significantly.

Threats to external validity relate to the generalizability of a technique. We agree that there might be rejection reasons that we could not identify. However, we analyze statistically significant samples of rejected edits by rollbacks from questions and answers. We thus believe that our manual investigation exposes all the main rejection reasons. Still, there is scope to analyze more samples to explore additional rejection reasons. However, suggested edits can be rejected by either rollback or expert review. Unfortunately, we could not collect samples of those edits rejected by expert reviews. We could not find any convenient way to collect such samples since their information is not readily available in the SO data dump. Thus, similar to existing literature, we consider the rejected edits by rollbacks \citep{Wang-SOEdit-TSE2018}. However, suggested edits are reviewed in SO by users with at least a 2K reputation score. In our manually analyzed dataset, 90.6\% (692 out of 764) of users who rolled back edits have a reputation score $\geq$ 2K. The remaining were self-rollback (i.e., rolled back by post owner). Therefore, our intuition is that the main reasons for rejections by expert reviews would be similar to our identified reasons. 



Our survey participants range from novice to experienced, mainly software developers and academicians (Table \ref{table:control-treatment-groups}). Such diversity in the survey participants offers validity and applicability to the survey findings. Furthermore, we ensure that control and treatment groups have participants with different professions and experience levels to mitigate individual bias.

We set a task list T1--T7 (Table \ref{table:edit-tasks}) and asked participants (control \& treatment group) to suggest edits to posts based on that list. However, tasks T1--T6 were set based on our identified rejection reasons. Thus, \texttt{EditEx} alerts the treatment group while suggesting T1--T6, but the control group does not get such alerts. As a result, the treatment group receives favor from \texttt{EditEx}, which might reduce their rejections. To mitigate this bias, we asked participants to suggest free-form edits (T7). From T7, we attempt to see the effectiveness of \texttt{EditEx} in assisting users in suggesting regular edits besides preventing common rejections. However, \texttt{EditEx} can support the SO edit system to prevent 12\% rejections. Such a finding confirms the effectiveness of \texttt{EditEx} in preventing not only common rejections but also rejections from regular edits.

\section{Deviations From Registered Report}
\label{sec:deviation}

This section discusses the deviations of this study from our registered report \citep{mondal2020rollbackRegisteredProtocol} and explains them.

\smallskip\noindent\textbf{Rollback Reasons \& Predictors.}
In our registered report, we wanted to see whether the addition/deletion of emotion influences edit rejection using EmoTxt \citep{calefato2017emotxt}. Our primary analysis found that emotion has almost no effect on edit rejection/acceptance. Furthermore, the overall accuracy of our rejected edit classifier improved only about 1\% if we consider emotion as a predictor. However, integrating a complex model to capture emotion and its deployment is costly. Therefore, it can affect the performance of our online tool \texttt{EditEx}. We thus discard emotion from this study.
On the other hand, we added \emph{reputation} as a predictor (Table \ref{table:features-of-predictor}). The reputation score estimates how much the community trusts a user \citep{anderson2012discovering, userReputation}. Therefore, we added \emph{Community Trust} as a rollback reason (which was absent in the registered report) (Table \ref{table:list-of-rollback-reasons}) that was estimated by reputation score. Moreover, users with lower reputation scores might violate the editing guidelines more than those with higher reputation scores. Note that violating edit guidelines is one of the causes of edit rejection. We thus consider reputation as a predictor that significantly improves the performance of our rejected edit classifier.
Another discrepancy to the registered report is that we included the \emph{Introducing Spam} rollback reason under \emph{Other}.

\smallskip\noindent\textbf{Manually Investigated Sample Size.}
In the registered report, we manually investigated 777 rollback edits (382 questions + 395 answers). The statistically significant sample size of rollback edits for both question and answer is 382. However, due to a programming problem, we randomly selected 395 samples from rejected answer revisions. We kept 395 samples in the registered report because - (1) we completed our analysis using 395 samples, and (2) 395 is more than the statistically significant sample size.  However, we later randomly selected 382 (among 395) samples (Section \ref{subsec:data-collection}) to equalize it with the statistically significant sample size and analyze them.

\smallskip\noindent\textbf{Recruitment of Participants.}
We planned to recruit 30 participants (15 for treatment + 15 for the control group) who edited at least 100 posts. After deploying our tool EditEx, we realized that EditEx could be helpful to both expert and novice SO users. Therefore, we relaxed our constraints. We recruit participants who edited any SO post to ensure familiarity with SO editing. However, we struggled to recruit 30 participants due to COVID-19 and the extensive nature of this user study. The study was extensive because each user had to do multiple edits of SO posts. Primarily, we planned to recruit participants using a snowball approach. To recruit more participants, we then extend our approach. Besides snowball, we attempt to recruit participants using an open circular. However, finally, we recruited 20 participants (10 for treatment + 10 for the control group) with different experience levels and diverse professions (Section \ref{subsec:participants}).

\smallskip\noindent\textbf{Number of Edits Per User.}
We planned to ask each participant to suggest ten edits. However, several participants (especially the control group) could not edit ten posts for three main challenges, as we discussed in Section \ref{sec:task-based-analysis}.

\smallskip\noindent\textbf{EditEx's Functionality of Highlighting Texts.}
We planned to include \texttt{EditEx}'s functionality to highlight texts that may cause rejection. However, the current version of \texttt{EditEx} cannot highlight texts. \texttt{EditEx} predicts the edit decisions (accepted/rejected) and alerts users with the potential rejection reasons if rejected. While highlighting texts could be helpful, we found that the \texttt{EditEx} tool with the basic features was usable and effective. Therefore, we leave the highlighting of texts in \texttt{EditEx} as a future extension.

\smallskip\noindent\textbf{NASA TLX Workload.}
We planned to estimate TLX effort as a task load by combining all the ratings provided by a participant in five dimensions in the TLX metrics. However, most existing studies estimate subjective workload by combining all the ratings in six dimensions \citep{cao2009nasa, hart1988development, noyes2007self, sharek2011useable, hart1986nasa}. Therefore, we also take ratings from each participant on six dimensions (mental demand, physical demand, temporal demand, effort, performance, and frustration). Finally, we estimate the cognitive workload (Fig. \ref{fig:nasa-tlx-workload}). However, we consider a scale of ten with one step size \citep{memarian2011work} to take ratings conveniently from participants for each dimension. It slightly sacrifices granularity in comparison to scale 100 with step size five. However, the results should not be affected much.

\smallskip\noindent\textbf{Unidentified Reasons.} 
Our rejection reason classifier cannot identify a couple of rejection reasons, such as partial acceptance and incorrect text/code change. We can extract added or removed text/code using appropriate HTML tags. However, partly accepted text/code cannot be separated from added text/code. Furthermore, it requires analyzing the future revisions to check partial acceptance, which is impractical. We did not find any patterns that can identify incorrect changes in code/texts. Identification of such reasons demands manual efforts.

\section{Related Work}
\label{sec:relatedwork}

We developed our tool \texttt{EditEx} to recommend fixes to suggested edits in SO so that SO users can avoid committing undesired edits that may lead to the rejection of the edits. As such, our research in this paper belongs to a broader area called `collaborative editing in social forums'. Major related work can broadly be divided into \textbf{Studies} of collaborative editing systems in crowd-sourced forums (see Section \ref{sec:related-studies}) and \textbf{Techniques} to suggest improvements to the editing system (see Section \ref{sec:related-techniques}). In addition, SO data are used extensively in SE research for various tasks (see Section \ref{sec:other-related-work-SO}), all of which could be potentially impacted by having low-quality data due to erroneous/inefficient edits. 

\subsection{Studies of Collaborative Editing Systems} 
\label{sec:related-studies}

Editing of content can improve the content quality. As such, it is intuitive that the social Q\&A forums offer to edit the post contents. Since social forums can be accessed by many users simultaneously, it is a cost-effective measure for the forums to support collaborative editing by allowing their users to do the editing. Indeed, studies 
show that collaborating editing in social forums and online collaborative knowledge-sharing portals (e.g., Wikipedia) can positively impact towards the improvement of shared contents~\citep{Li-TradeOffCollaborativeEditingWikipedia-CSCW2015,Kittur-HarnessingWisdomCrowdWikipedia-CSCW2008}. The nature of the collaborative editing can be similar across the social forums (e.g., Q\&A site) and knowledge portals (e.g., Wikipedia). The research of Li et al.~\citep{Li-TradeOffCollaborativeEditingWikipedia-CSCW2015} looked at the adoption of Wikipedia-style collaborative editing into a Q\&A site like SO. They found that users with good edits are rewarded with positive votes by other users. They analyzed five years of historical editing data from SO and found that substantive edits from other users can increase the number of positive votes by 18\% for the questions and 119\% for answers. This reward can be beneficial for a user who does the edit because the edit may only offer at most 5\% improvement over the original post (i.e., the user can be rewarded with mindful but low-cost editing efforts). 
Indeed, the SO reward system can serve as an added influence to the users to suggest edits. A recent study by Wang et al.~\citep{Wang-SOEdit-TSE2018} in SO found that users are motivated to edit more when they are closer to getting 
a badge. 

Overall, both Wang et al.~\citep{Wang-SOEdit-TSE2018} and Li et al.~\citep{Li-TradeOffCollaborativeEditingWikipedia-CSCW2015} conclude that offering incentives as reputation scores is useful to improve post quality within a collaborative editing platform like SO.
This finding was also observed in other collaborative editing platforms like webcasts~\citep{Munteanu-CollaborativeEditingWebcasts-CHI2008} and Wikipedia~\citep{Kittur-HarnessingWisdomCrowdWikipedia-CSCW2008}. 
Munteanu et al.~\citep{Munteanu-CollaborativeEditingWebcasts-CHI2008} tested the effectiveness of engaged users in collaborating in a wiki-like webcast platform to edit/correct transcripts that are produced from webcasts through 
an automated speech recognition system. Collaborative editing can be a cost-effective but useful means to improve the quality of the ASR (Automated Speech Recognition) system in webcasts because ASR systems can have an average error rate of 45\% - above the accepted threshold of 25\%. The field study carried out by the authors in a real lecture environment found that using students to edit the webcast transcript was useful in reducing the error rate. The editing was supported via a webcast extension that engages users to collaborate in a wiki-like manner. 
Kittur et al.~\citep{Kittur-HarnessingWisdomCrowdWikipedia-CSCW2008} find that the increase in the number of editors does not guarantee the quality of the articles on Wikipedia. 

The quality of the question is important to get an answer: lack of clarity, relatedness, and reproducibility of the problem, as well as the too short question, could dissuade developers from answering the question~\citep{Asaduzzaman-FindingAnswersToUnansweredQuestions-MSR2013,Mondal-SOIssueReproducability-MSR2019}. The reputation 
and past activity of an asker could also factor into the likelihood of a question getting resolved~\citep{Masud-InsightUnresolvedQuestionsSO-MSR2015}. As such 
factors of good questions are investigated, e.g., code to text ratio, etc.~\citep{Calefato-HowToAskForTechnicalHelp-IST2018,Duijn-QualityQuestionsNeedQualityCode-MSR2015}. 
However, depending on the platforms and user characteristics, these factors can vary~\citep{Hudson-TriggerForClarificationRequestsForums-VLHCC2015}. 
As such, it is important to detect content quality automatically~\citep{Ponzanelli-ClassifyQualityForumQuestion-QSIC2014,Ponzanelli-ImprovingSOPosts-ICSME2014, Ya-DetectHighQualityPosts-JIS2015}. Wang et al.~\citep{Wang-SOEdit-TSE2018} found that users who make more edits in a short time are likely to get more edits rejected. Thus bad edits can harm the content quality.

Our research on SO rollback edits initially started in 2019 to better understand the edit rejection reasons as reported by Wang et al.~\citep{Wang-SOEdit-TSE2018}. Through our qualitative analysis of SO posts, we also found all the edit rejection reasons reported by Wang et al.~\citep{Wang-SOEdit-TSE2018}. In addition, we found four more edit rejection reasons. We report the edit rejection reasons in Section \ref{sec:rejection-catalog} of this paper. While the above papers, including Wang et al.~\citep{Wang-SOEdit-TSE2018}, focus on analyzing editing mechanisms in collaborative platforms based on empirical studies, our paper focuses on developing techniques to automatically suggest fixes to suggested edits so that the edits will not be rejected upon submission. As such, our developed tool \texttt{EditEx} can further contribute to supporting the content quality in social forums by assisting users with guidance on improving the quality of their suggested contents. Thus, our paper offers complementary viewpoints to the above studies by offering tools and techniques that can facilitate improved edit content in a social Q\&A site like SO.

\subsection{Techniques to Develop to Improve Collaborative Editing Systems}
\label{sec:related-techniques} 

Collaborative editing systems are common in Wikipedia ~\citep{Li-TradeOffCollaborativeEditingWikipedia-CSCW2015,Kittur-HarnessingWisdomCrowdWikipedia-CSCW2008}, 
GitHub code editing~\citep{Dabbish-SocialCodingGithubTransparency-CSCW2012}, webcasts~\citep{Munteanu-CollaborativeEditingWebcasts-CHI2008}, 
scientific contents~\citep{Lowry-TaxonomyCollaborativeEditingForEmpiricalResearch-JBC2005,
Calvo-CollaborativeAndWritingToolsCloud-JBC2005}, and so on. Compared to substantial research on conducting studies on existing collaborative editing systems, we are not aware of much research that focuses on developing tools and techniques to improve the systems. This is perhaps due to the fact that currently available collaborative platforms like Wikipedia seem to work well and are hugely popular. In all these platforms, the focus of collaborative editing is to improve the quality of the shared content based on 
user engagement~\citep{Agichtein-FindingHighQualityContents-WSDM2008}. 

Chen et al.~\citep{Chen-DeepLearningCollaborativeEdit-CSCW2017} observed that most of the edits in SO are small sentence edits. While developing their SOTorrent database, Baltes et al.~\citep{Baltes-SOTorrentEvolution-MSR2018} also observed that majority of edits in SO are relatively small. In a follow-up study, 
Chen et al.~\citep{Chen-DataDrivenPolicyPostQuality-CSCW2018} predicted whether a post needs to be edited. Their approach is based on the concept of `proactive policy assurance', which assures that a modification to a suggested edit will satisfy the current `reactive policy assurance' in SO, which accepts/rejects based on the matching of exiting editing policy after an edit is submitted (i.e., reactive). 
They developed a deep-learning-based policy assurance tool to recommend post owners or other users' potential mid-level edits to given post content. The deep learning model is a CNN (Convolutional Neural Network). In a large-scale experiment, they find that the tool offers good precision, recall, and F1-score (at least 0.7) while suggesting mid-level edits. 

As we noted in Section \ref{sec:related-studies}, our research of this paper started in 2019 to gain hands-on experience on the edit rejection reasons observed by Wang et al.~\citep{Wang-SOEdit-TSE2018}. Our initial exploration led to an expansion of the edit rejection reasons and to the submission of a registered protocol report in 2020 \citep{mondal2020rollbackRegisteredProtocol}. In the registered protocol report, we outlined our vision of this paper by offering to develop machine learning model to automatically detect the edit rejection reasons and to build our \texttt{EditEx} tool that can offer proactive guidance to fix suggested edits. While working on this paper, we observed that some edit rejection reasons could be present both in accepted and rejected edits, resulting in \textit{inconsistencies} in the editing acceptance/rejection process. We reported a catalog of such inconsistencies in our MSR 2021 paper \citep{mondal2021rollback}. In our MSR 2021 paper, we also report several rule-based tools that we developed to detect inconsistencies in SO edits automatically. While developing our \texttt{EditEx} tool in this paper, we purposefully did not consider those inconsistencies, given those were not outlined in our 2020 registered protocol report \citep{mondal2020rollbackRegisteredProtocol}. We note that an immediate extension of \texttt{EditEx} could investigate whether and how the inclusion of the inconsistencies into the rejection prediction models and the \texttt{EditEx} tool could make the overall editing process more effective for the SO users. We leave it as our immediate future work.

\subsection{Other SE Research using SO Data}
\label{sec:other-related-work-SO}

Several studies have been conducted to study developer discussions on different crowd-shared developer platforms, including SO. Seaman et al. studied developer discussion on inspection meetings~\cite{seaman1998communication}. Reiner et al. used content analysis to study developer discussions on software processes~\cite{rainer2003persuading}. Gottipati et al. study relevant answers in 3 software forums, Dzone, Tips, and Oracle forums~\cite{gottipati2011finding}. Several studies has focused on discussions on microblogs, such as, Twitter~\cite{tian2012does,prasetyo2012automatic,wang2013microblogging}, and chat communities, such as, HipChat~\cite{alkadhi2017rationale}, IRC messages~\cite{alkadhi2018developers,shihab2009studying}, and Slack~\cite{chatterjee2019exploratory}. Recently, SO Q\&A forums have been subject to a number of papers to study various aspects of software development, such as what developers are discussing in general~\cite{Barua-StackoverflowTopics-ESE2012},
or about a particular aspect, e.g., concurrency~\cite{Ahmed-ConcurrencyTopic-ESEM2018}, big data~\cite{Bagherzadeh2019}, chatbot development~\cite{abdellatifchallenges}. 

Several studies has been conducted to study developer sentiments on online discussions (e.g., SO data)~\cite{guzman2014sentiment,murgia2014developers,ortu2015bullies,novielli2014towards,Uddin-OpinionValue-TSE2019,Uddin-OpinerReviewAlgo-ASE2017,Uddin-OpinerReviewToolDemo-ASE2017,Uddin-OpinionSurvey-TSE2019,Chakraborty-NewLangSupportSO-IST2021,Uddin-OpinerAPIUsageScenario-TOSEM2020,Uddin-OpinerAPIUsageScenario-IST2020,Lin-Opinion-TOSEM2022}. Guzman et. al. applied sentiment analysis on code comments~\cite{guzman2014sentiment}. Islam and Zibran study emotional variations in commit messages~\cite{islam2016towards}. Garcia et al. ~\cite{garcia2013role} study the emotions of developers in the Gentoo community. Guzman and Bruegge studied developer sentiments on mailing lists~\cite{guzman2013towards}. Novielli et al. conduct sentiment analysis on SO and Github discussions~\cite{novielli2015challenges}. Many of these studies use automated sentiment analysis tools, which are found to provide contradictory results in software engineering research~\cite{jongeling2017negative}. 

All the above research works using SO data could be benefited from improved data quality offered by collaborative editing in SO. As such, our tool \texttt{EditEx}, once adopted by the SO users, can help the SO users as well as the SO-based research community with better quality data.

\section{Conclusion}
\label{sec:conclusion}

SO has become an essential online resource with millions of programming-related problems and solutions. However, the quality of the shared knowledge is vital for the growth and success of SO. To promote quality, SO introduces an edit system so that users can suggest an improvement to posts. Unfortunately, numerous suggested edits are rejected due to either undesired changes of posts or violating edit guidelines. Such a scenario not only hurts the quality of content but also frustrates and demotivates users. We conducted a qualitative analysis of 764 (382 questions + 382 answers) rejected edits by rollbacks and identified 19 rejection reasons. We then extract 15 texts and user-based features to automatically capture those reasons and develop four machine learning models using them. Our best-performing model can predict rejected edits with about 70\% accuracy, and the rejection reason classifier can identify the potential rejection reasons with 67\% accuracy.
We also introduced an online tool named \texttt{EditEx} that can be integrated with the SO edit system. It analyzes the edits, predicts whether they will be rejected, and suggests users with the potential rejection reasons.
We conduct a survey to assess \texttt{EditEx} and SO edit system. According to survey results, the participants find reasons for rejection identified by \texttt{EditEx} \emph{influential}. Moreover, \texttt{EditEx} can support the SO edit system to prevent 49\% rejections, including the commonly rejected reasons. Such a statistic is 12\% when users suggest regular free-form edits. Moreover, our tool significantly decreases the subjective workload and increases participants' confidence in suggesting edits.

\begin{acknowledgements}
This research is supported in part by the Natural Sciences and Engineering Research Council of Canada (NSERC) Discovery grants, and by an NSERC Collaborative Research and Training Experience  (CREATE) grant, and by two Canada First Research Excellence Fund (CFREF) grants coordinated by the Global Institute for Food Security (GIFS) and the Global Institute for Water Security (GIWS).
\end{acknowledgements}

\bibliographystyle{spbasic}

\bibliography{reference.bib}

%
%

\end{document}